\documentclass[aps,twocolumn,pra,showpacs,a4paper,floatfix,preprintnumbers,amsmath,amssymb]{revtex4}

\usepackage{exscale}
\usepackage{graphicx}
\usepackage{amsmath}
\usepackage{latexsym}
\usepackage{amsfonts}
\usepackage{amssymb}
\usepackage{amscd}
\usepackage{bm}           
\usepackage{bbm}
\usepackage[english]{babel}
\usepackage{latexsym}
\usepackage{graphics}
\usepackage{epsfig}
\usepackage{color}
\usepackage{mathrsfs}
\usepackage{hyperref}
\hypersetup{
colorlinks=true,
citecolor=blue,
linkcolor=red,
urlcolor=black
}

\include{./journalsVaps}

\newcommand{\vect}[1]{\bm{#1}}

\newcommand{\bra}[1]{\ensuremath{\langle#1|}}

\newcommand{\ket}[1]{\ensuremath{|#1\rangle}}

\newcommand{\ketbra}[1]{\ensuremath{| #1 \rangle \langle #1 |}}

\newcommand{\Eins}{\ensuremath{\mathbbm 1}}

\newcommand{\BE}{\begin{equation}}
\newcommand{\EE}{\end{equation}}
\newcommand{\be}{\begin{equation}}
\newcommand{\ee}{\end{equation}}
\newcommand{\bea}{\begin{eqnarray}}
\newcommand{\eea}{\end{eqnarray}}
\newcommand{\beq}{\begin{eqnarray}}
\newcommand{\eeq}{\end{eqnarray}}
\newcommand{\bean}{\begin{eqnarray*}}
\newcommand{\eean}{\end{eqnarray*}}
\newcommand{\kommentar}[1]{}

\newcommand{\mean}[1]{\ensuremath{\langle #1 \rangle}}

\newcommand{\proj}[1]{\ketbra{#1}}
\newcommand{\tr}{{\rm Tr}}

\newcommand{\bc}{\begin{center}}
\newcommand{\ec}{\end{center}}

\newcommand{\ud}{\mathrm{d}}
\newcommand{\ie}{{\it i.e.}}

\begin{document}

\title{Phase sensitivity bounds for two-mode interferometers}

\author{Luca Pezz\`e$^{1}$, Philipp Hyllus$^{2}$ and Augusto Smerzi$^{1}$}

\affiliation{
$^1$INO-CNR and LENS, Largo Enrico Fermi 6, I-50125 Firenze, Italy\\
$^2$Department of Theoretical Physics, University of the Basque Country EHU/UPV, P.O. Box 644, E-48080 Bilbao, Spain
}


\date{\today}

\begin{abstract}
We provide general bounds of phase estimation sensitivity in linear two-mode interferometers.
We consider probe states with a fluctuating total number of particles. 
With incoherent mixtures of state with different total number of particles, particle entanglement is 
necessary but not sufficient to overcome the shot noise limit.
The highest possible phase estimation sensitivity, the Heisenberg limit, is established 
under general unbiased properties of the estimator. 
When coherences can be created, manipulated and detected,
a phase sensitivity bound can only be set in the central limit, with 
a sufficiently large repetition of the interferometric measurement. 
\end{abstract}

\maketitle


\section{Introduction}     

The problem of determining the ultimate phase sensitivity (often tagged as the ``Heisenberg limit'') 
of linear interferometers has long puzzled the field
\cite{Caves_1981,SummyOPTCOMM1990,HradilPRA1995,HradilQO1992,HallJMO1993,Ou_1996,Shapiro_1989, Shapiro_1991,Bondurant_1984, Schleich_1990,Braunstein_1992,Wineland_1994,Holland_1993,YurkePRA1986}
and still arises controversies 
\cite{Durkin_2007,Monras_2006,BenattiPRA2013, ZwierzPRL2010, JarzynaPRA2012, Giovannetti_2006, Hyllus_2010, Pezze_2009, 
Rivas_2011, PezzePRA2013, Joo_2011, HayashiPI2011, GLM_preprint2011, Tsang_preprint2011, HallPRA2012, BerryPRA2012, HallNJP2012, Hofmann_2009, AnisimovPRL10, ZhangJPA2012, GaoJPA2012, Giovannetti_2012}.
The recent revival of interest is triggered by the current impressive experimental 
efforts in the direction of quantum phase estimation with ions \cite{exp_ions}, cold atoms \cite{exp_coldatoms}, 
Bose-Einstein condensates \cite{exp_BEC} and photons \cite{exp_photons}, 
including possible applications to large-scale gravitational wave detectors \cite{Schnabel_2010}.
Beside the technological applications, the problem is closely related to
fundamental questions of quantum information, most prominently, 
regarding the role played by quantum correlations.
In particular, the phase sensitivity of a linear two-mode interferometer 
depends on the entanglement between particles (qubits) in the 
input (or ``probe'') state \cite{Giovannetti_2006, Hyllus_2010, Pezze_2009}.
It is widely accepted \cite{Giovannetti_2006,Pezze_2009} that, when the number 
of qubits in the input state of is fixed, and equal to $N$
[so that the mean square particle-number fluctuations $(\Delta \hat{N})^2=0$], 
there are two important bounds in the uncertainty of {\it unbiased} phase estimation.
The shot noise limit, 
\be \label{SN_fixed}
\Delta \theta_{\rm{SN}} = \frac{1}{\sqrt{m\, N}}, \qquad \mathrm{for\,\,} (\Delta \hat N)^2 = 0,
\ee
is the maximum sensitivity achievable with probe states containing 
only classical correlations among particles.
The factor $m$ accounts for the number of independent repetitions of the measurement. 
This bound is not fundamental. It can be surpassed by preparing the $N$ particles of the probe
in a proper entangled state.
The fundamental (Heisenberg) limit is given by
\be \label{HL_fixed}
\Delta \theta_{\rm{HL}} = \frac{1}{\sqrt{m}\, N}, \qquad \mathrm{for\,\,} (\Delta \hat N)^2 = 0,
\ee
and it is saturated by maximally entangled (NOON) states.

It should be noticed that most of the theoretical investigations
have been  developed in the context of systems having a fixed, known, total number of particles $N$.
However, many experiments are performed in presence of finite fluctuations 
$(\Delta \hat N)^2 > 0$.
The consequences in the phase sensitivity of classical and quantum fluctuations of the number of particles
entering the interferometer have not been yet investigated in great depth.
In this case, indeed, the existence and discovery of the phase uncertainty bounds can be critically 
complicated by the presence of coherences between different total number of particles 
in the probe state and/or the output measurement \cite{Hyllus_2010,Hofmann_2009}.
However such quantum coherences do not play any role in two experimentally relevant cases:
{\it i)} in the presence of superselection rules, which are especially relevant for 
massive particles and forbid the existence of number coherences
in the probe state; 
{\it ii)} when the phase shift is estimated by measuring an arbitrary function of the 
number of particles in the output state of the interferometer,
{\it e.g.} when the total number of particles is post-selected by the measurement apparatus.
The point {\it (ii)} is actually an ubiquitous condition in current atomic and optical experiments.
Indeed, all known phase estimation protocols implemented experimentally are realised by measuring particle numbers.

In the absence of number coherences, or when coherences are present but irrelevant because of {\it (ii)}, 
we can define a state as separable if it is separable in each subspace of a fixed number of particles \cite{Hyllus_2010}.
A state is entangled if it is entangled in at least one subspace of fixed number of particles. 
With separable states, the maximum sensitivity of {\it unbiased} phase estimators is bounded by the shot-noise
\be 
\Delta \theta_{\rm{SN}} = \frac{1}{\sqrt{m\, \langle \hat N \rangle}},
\qquad \mathrm{for\,\,} (\Delta \hat N)^2 > 0,
\ee
while with entangled states the relevant bound, the Heisenberg limit, is given by  \cite{Hyllus_2010}
\be \label{HL_nonfixed}
\Delta \theta_{\rm{HL}} = 
\max\left[ \frac{1}{\sqrt{m \langle \hat N^2 \rangle}}, \frac{1}{m \langle \hat N \rangle} \right],
\quad \mathrm{for\,\,} (\Delta \hat N)^2 > 0.
\ee
We point out that Eq.~(\ref{HL_nonfixed}) cannot be obtained 
from Eq.~(\ref{HL_fixed}) by simply replacing $N$ with $\langle \hat N \rangle$. 
On the other hand, Eq.~(\ref{HL_nonfixed}) reduces to Eq.~(\ref{HL_fixed})
when number fluctuations vanish, $\mean{\hat N^2} = \mean{\hat N}^2$.
An example of phase estimation saturating the scaling $1/m \langle \hat N \rangle$ is obtained  
with the coherent$\otimes$squeezed-vacuum state \cite{PezzePRL2008}.

When the probe state {\it and} the output measurement contain
number coherences, the situation becomes more involved.
It is still possible to show that Eq.~(\ref{HL_nonfixed}) holds in the central limit ($m \gg 1$), at least.
Outside the central limit, it is possible to prove that the highest 
phase sensitivity is bounded by 
\be \label{HL}
\Delta \theta_{\rm{QCR}} = \frac{1}{\sqrt{m \langle \hat N^2 \rangle}}.
\ee 
The crucial point is that the fluctuations $\mean{\hat N^2}$ can be made arbitrarily
large even with a finite $\mean{\hat N}$. In general, 
no lower bound can be settled in this case: 
$\Delta \theta \geq 0$, and it can be saturated with {\it finite} resources ($m < \infty$, $\mean{\hat N}< \infty$)
if an unbiased estimator exists.
Outside the central limit, ({\it i.e.} for a small number of measurements) we cannot rule out 
the existence of opportune unbiased estimators which can saturate Eq. (\ref{HL}).

This manuscript extends and investigate in details the results and concepts introduced in 
Ref.~\cite{Hyllus_2010}.
In Sec.~\ref{Basic} we review the theory of multiparameter estimation
with special emphasis on two-mode linear transformations.
This allows us to introduce the useful concept of (quantum) Fisher 
information and the Cram\'er-Rao bound.
We show that two mode phase estimation involves, 
in general, operations which belong to the U(2)  transformation group. 
When number coherences in the probe state and/or in output measurement observables are not present,
the only allowed operations are described by SU(2) group. 
In Secs.~\ref{BoundsFisherSU2} and \ref{BoundsFisherU2} we give bounds 
on the quantum Fisher information depending whether or not the probe state contains 
number coherences.
In the latter case, we set an ultimate bound that can be 
reached by separable states of a fluctuating number of particles. 
Finally, in Sec.~\ref{SecHL} we discuss the Heisenberg limit, Eq.~(\ref{HL_nonfixed}),
and under which conditions it holds.

This manuscript focuses on the ideal noiseless case. 
It is worth pointing out that decoherence can strongly affect 
the achievable phase uncertainty bounds. 
For several relevant noise models in quantum metrology,
as, for instance, particles losses, correlated or uncorrelated phase noise, 
phase uncertainty bounds have been derived 
\cite{HuelgaPRL1997, ShajiPRA2007, EscherNATPHYS2011, RafalNATCOMM2012, DornerNJP2012, LandiniNJP2014}.


\section{Basic concepts}
\label{Basic}

In the (multi-)phase estimation problem, a probe state $\hat \rho$ undergoes 
a transformation which depends on the unknown vector parameter $\vect{\theta}$.
The phase shift is estimated from measurements of the transformed state $\hat \rho_{\rm{out}}(\vect{\theta})$.
The protocol is repeated $m$ times by preparing identical copies of $\hat \rho$
and performing identical transformations and measurement.
The most general measurement scenario is a positive-operator valued measure (POVM), \ie~a set of non-negative 
Hermitian operators $\{\hat E(\varepsilon)\}_\varepsilon$ parametrized by $\varepsilon$
and satisfying the completeness relation $\int \ud \varepsilon \hat E(\varepsilon) = \Eins$ \cite{Helstrom_book}.
The label $\varepsilon$ indicates the possible outcome of a measurement
which can be continuous (as here), discrete or multivariate.
Each outcome $\varepsilon$ is characterized by a probability 
$P(\varepsilon \vert \vect{\theta}) = \tr[ \hat E(\varepsilon) \hat \rho_{\rm{out}}(\vect{\theta}) ]$, 
conditioned by the true value of the parameters.
The positivity and Hermiticity of $\{\hat E(\varepsilon)\}_\varepsilon$ guarantee that 
$P(\varepsilon \vert \vect{\theta})$ are real and nonnegative, the completeness guarantees that 
$\int \ud \varepsilon P(\varepsilon \vert \vect{\theta})=1$.
The aim of this section is to settle the general theory of phase estimation for
two-mode interferometers.


\subsection{Probe state}

A generic probe state with fluctuating total number of particles
can be written as 
\be \label{eq:coherent}
\hat\rho_{\rm coh}=\sum_k p_k \proj{\psi_k}
\ee
with $p_k>0$ and $\sum_k p_k=1$, where 
\be
\ket{\psi_k}=\sum_N \sqrt{Q_{N,k}} \, \ket{\psi_{N,k}}
\ee
is a coherent superposition of states $\ket{\psi_{N,k}}$ with different number of particles.
The coefficients $Q_{N,k}$ are complex numbers and the 
normalization condition $\bra{\psi_k}\psi_k\rangle=1$ implies $\sum_N \vert Q_{N,k} \vert =1$.
It is generally believed that quantum coherences
between states of different numbers of particles do not play any observable
role because of the existence of superselection rules (SSRs) for the total number
of particles \cite{WickPR52,Bartlett_2007}. 
In the presence of SSRs the only physically meaningful states are those 
which commute with the number of particles operator, 
\be \label{commrhoN}
[\hat \rho, \hat N] = 0.
\ee
A state satisfies this condition if and only if \cite{nota:commutatorN} 
it can be written as the incoherent mixture 
\be \label{eq:incoherent}
\hat\rho_\mathrm{inc}=\sum_N  Q_N  \, \hat \rho^{(N)},
\ee 
where $\hat\rho^{(N)}$ is a normalized ($\tr[\hat\rho^{(N)}]=1$) state, 
$Q_N  = \tr[\hat \pi_N \hat\rho \hat \pi_N]$ are positive numbers satisfying $\sum_N  Q_N =1$
and $\hat \pi_N$ are projectors on the fixed-$N$ subspace.
The existence of SSRs is the consequence of the lack of a 
phase reference frame (RF) \cite{Bartlett_2007}. 
However, the possibility that a suitable RF can be established 
in principle cannot be excluded \cite{Bartlett_2007}.
If SSRs are lifted, then coherent superpositions
of states with different numbers of particles become physically relevant.


\subsection{separability and multiparticle entanglement}
\label{SubSecEnt}

A crucial property of the probe state is particle entanglement. 
A state of $N$ particle is called separable if it can be written as
a convex sum of product states~\cite{WernerPRA89, GuhnePHYSREP2009},
\be \label{rhosep}
\hat{\rho}^{(N)}_{\rm sep}=\sum_k P_{k} \proj{\phi_{k}^{(1)}} \otimes \cdots \otimes \proj{\phi_{k}^{(N)}},
\ee
where $\ket{\phi_{k}^{(i)}}$ is the state of the $i$th particle.
A state is (multiparticle) entangled if it is not separable.
One can further consider the case where only a 
fraction of the $N$ particles are in an entangled state and 
classify multiparticle entangled states following Refs.~\cite{SeevinckPRA01, GuehneNJP05, ChenPRA05, SoerensenPRL01, GuhnePHYSREP2009}. 
A pure state of $N$ particles is $k$-producible if it can be written
as $\ket{\psi_{k-{\rm prod}}} = \otimes_{l=1}^M \ket{\psi_l}$,
where $\ket{\psi_l}$ is a state of $N_l \leq k$
particles, witht $\sum_{l=1}^M N_l=N$.
A state is $k$-particle entangled if 
it is $k$-producible but not $(k-1)$-producible.
Therefore, a $k$-particle entangled state can be written 
as $\ket{\psi_{k-{\rm ent}}} = \otimes_{l=1}^M \ket{\psi_l}$
where the product contains at least one state $\ket{\psi_l}$ of $N_l=k$ particles which does not factorize.
A mixed state is $k$-producible if it can be written as a mixture of 
$(k_l\le k)$-producible pure states, {\em {\it i.e.}},
$\rho_{k-{\rm prod}}=\sum_l p_l \proj{\psi_{k_l-{\rm prod}}}$, where $k_l\le k$ for all $l$.
Again, it is $k$-particle entangled if it is $k$-producible but not $(k-1)$-producible.
Notice that, formally, a separable state is $1$-producible,
and that a decomposition of a $k<N$-particle entangled state of $N$ particles
may contain states where different sets of particles are entangled.

We here extend the definition of separability/entanglement to states of a fluctuating number of particles.
An incoherent mixture (\ref{eq:incoherent}) is defined as separable if it can be written as \cite{Hyllus_2010}
\be\label{sep}
\hat\rho_{\rm sep}=\sum_{N} Q_N \hat\rho^{(N)}_{\rm sep},
\ee
where $\hat{\rho}^{(N)}_{\rm sep}$ is a separable state of $N$ particles, see Eq.~(\ref{rhosep}).
States which are not separable according to this definition are entangled.
Similarly, an incoherent mixture is $k$-producible if \cite{HyllusPRA2012}
\be \label{rho_inc}
\rho_{k-{\rm prod}}=\sum_N Q_N \rho^{(N)}_{k-{\rm prod}},
\ee
where $\rho^{(N)}_{k-{\rm prod}}$ is a $k$-producible state of $N$ particles. 
A state $\hat\rho_{\rm coh}$ with number coherences (\ref{eq:coherent}) will be called separable 
if it is separable in every fixed-$N$ subspace~\cite{Hyllus_2010}, 
{\em {\it i.e.}} if the incoherent mixture $\sum_N \hat \pi_N \hat\rho_{\rm coh} \hat \pi_N$,
obtained from $\hat\rho_{\rm coh}$ by projecting over fixed-$N$ subspaces,
has the form of Eq.~(\ref{sep}). 
Analogously, a state will be called $k$-producible if the projection on each fixed-$N$
subspace has the form of Eq.~(\ref{rho_inc}).

\begin{figure*}[t!]
\begin{center}
\includegraphics[scale=0.4]{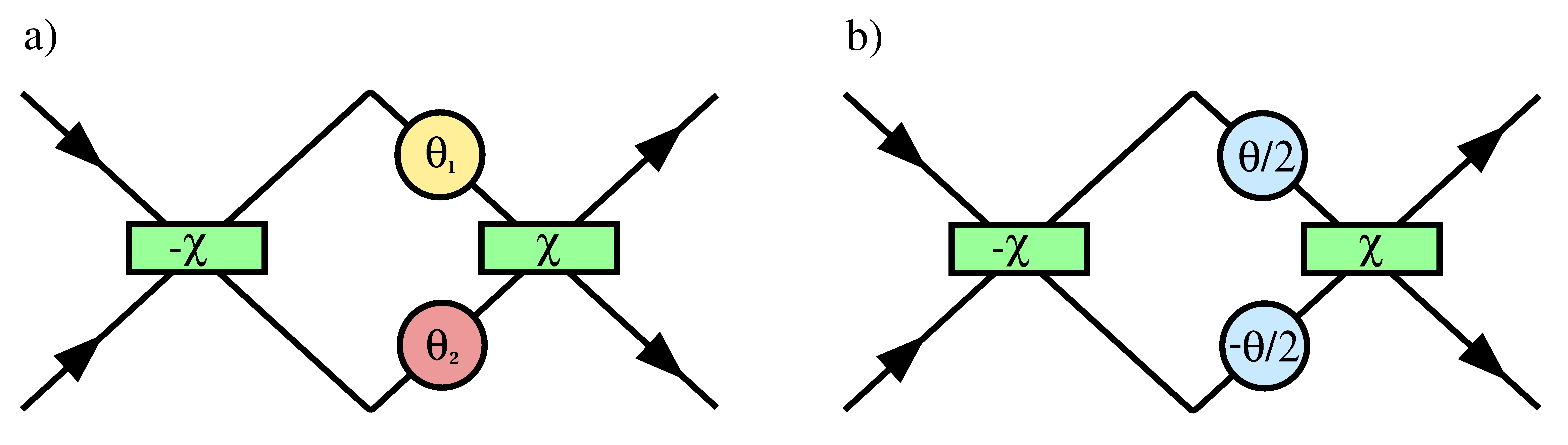}
\end{center}
\caption{ (color online) Schematic representation of U(2) (a) and SU(2) (b) interferometers.
In the U(2) case, the general transformation is given by Eq.~(\ref{TMtransformU2}),
where $a_1$ and $a_2$ are the modes inside the interferometer and 
the green squares represent the transformation $e^{\pm i \chi \hat{J}_{\vect{s}}}$.
In the SU(2) case the only phase to estimate is the relative phase shift $\theta$
among the $a_1$ and $a_2$ modes and the 
general transformation is given by Eq.~(\ref{TMtransformSU2}).}  \label{fig:Interf}
\end{figure*}

\subsection{Two-mode transformations}
\label{Sec:TwoMode}

In the following we will focus on linear transformations involving two modes. 
These includes a large class of optical and atomic passive devices, 
including the beam-splitter, Mach-Zehnder and Ramsey interferometers.
Most of the current prototype phase estimation experiments 
\cite{exp_ions,exp_coldatoms,exp_photons,exp_BEC} are well described by a 
two-mode approximation.

Denoting by $\hat{a}_1$ and $\hat{a}_2$ ($\hat{b}_1$ and $\hat{b}_2$)
are input (output) mode annihilation operators, 
we can write 
\be 
\left[
\begin{array}{c}
\hat{b}_1 \\
\hat{b}_2 \\
\end{array} 
\right] 
=
\mathbf{U}\, 
\left[
\begin{array}{c}
\hat{a}_1 \\
\hat{a}_2 \\
\end{array} 
\right], 
\ee
where $\mathbf{U}$ is a $2 \times 2$ matrix~\cite{CamposPRA1989, Leonhardt_1993, YurkePRA1986}.
By imposing the conservation of the total number of particles, 
$\hat{a}_1^{\dag}\hat{a}_1+\hat{a}_2^{\dag}\hat{a}_2 = \hat{b}_1^{\dag}\hat{b}_1+\hat{b}_2^{\dag}\hat{b}_2$,
we obtain that $\mathbf{U}$
can be explicitly written as
\be \label{ModeMatrix}
\mathbf{U}=
e^{-i\phi_0} 
\left[
\begin{array}{cc}
 e^{-i \phi_t} \cos \frac{\vartheta}{2} & -e^{-i \phi_r} \sin \frac{\vartheta}{2} \\
e^{i \phi_r} \sin \frac{\vartheta}{2} & e^{i \phi_t} \cos \frac{\vartheta}{2} \\ 
\end{array} 
\right]. 
\ee
The matrix Eq.~(\ref{ModeMatrix}) is unitary, 
preserves bosonic and fermonic commutation relations between 
the input/output mode operators and 
its determinant is equal to $e^{-2i\phi_0}$.
The most general two mode transformation thus
belongs to the U(2)=U(1)$\times$SU(2) group (unitary matrices with determinant 
$\vert \rm{det}\mathbf{U} \vert =1$).
The coefficients $\vartheta$ is physically related to transmittance $t=\cos^2 \vartheta/2$
and reflectance $r=\sin^2 \vartheta/2$ of the transformation (\ref{ModeMatrix}), 
$\phi_t$ and $\phi_r$ being the corresponding phases.
The lossless nature of Eq.~(\ref{ModeMatrix}) is guaranteed by $t+r=1.$ 

Using the Jordan-Schwinger representation of angular momentum systems 
in terms of mode operators~\cite{SchwingerBOOK}, it is possible 
to find the operator $\hat{\mathrm{U}}$
corresponding to the matrix~(\ref{ModeMatrix}).
In other words, $\hat{b}_i = \hat{\mathrm{U}}^\dag \hat{a}_i \hat{\mathrm{U}}$ for $i=1,2$
is the transformation of mode operators (Heisenberg picture) and 
$\hat \rho_{\rm{out}}= \hat{\mathrm{U}} \hat \rho \hat{\mathrm{U}}^{\dag}$,
$\vert \psi_{\rm{out}} \rangle = \hat{\mathrm{U}} \vert \psi_{\rm{in}} \rangle$,
is the equivalent transformation of statistical mixtures and quantum states, respectively (Schr\"odinger picture).
One finds~\cite{CamposPRA1989, YurkePRA1986}
\be \label{ModeOperator}
\hat{\mathrm{U}}(\phi_0,\theta) = e^{-i \phi_0 \hat{N}} \, e^{-i\theta \hat{J}_{\vect{n}}},
\ee 
where 
\be
\hat{N} = \hat{a}^{\dag}_1 \hat{a}_1 + \hat{a}^{\dag}_2 \hat{a}_2 \nonumber
\ee
is the number of particle operator, 
$\hat{J}_{\vect{n}} = \alpha \hat{J}_{\vect{x}} + \beta \hat{J}_{\vect{y}} + \gamma \hat{J}_{\vect{z}}$
(where $\alpha$, $\beta$ and $\gamma$ are the coordinates 
of the vector $\vect{n}$ in the Bloch sphere and satisfy $\alpha^2 + \beta^2 + \gamma^2=1$), and
\beq 
\hat{J}_{\vect{x}} &=& \frac{\hat{a}^{\dag}_1 \hat{a}_2 + \hat{a}^{\dag}_2 \hat{a}_1}{2}, \nonumber \\
\hat{J}_{\vect{y}} &=& \frac{\hat{a}^{\dag}_1 \hat{a}_2 - \hat{a}^{\dag}_2 \hat{a}_1}{2i}, \nonumber \\
\hat{J}_{\vect{z}} &=& \frac{\hat{a}^{\dag}_1 \hat{a}_1 - \hat{a}^{\dag}_2 \hat{a}_2}{2}, \nonumber 
\eeq
are spin operators.
The exact relation between the parameters of the matrix $\mathbf{U}$ 
[$\phi_\tau$, $\phi_\rho$ and $\vartheta$ in Eq.~(\ref{ModeMatrix})] and the parameters 
of the operator $\hat U$ [$\theta$, $\alpha$, $\beta$ and $\gamma$ in Eq.~(\ref{ModeOperator})]
is given in Appendix~A.
The operators $\hat{J}_{\vect{x}}$, $\hat{J}_{\vect{y}}$ and $\hat{J}_{\vect{z}}$
satisfy the angular momentum commutation relations.
Notice that the pseudo-spin operators commute with 
the total number of particles, $[\hat{J}_{\vect{k}},N]=0$ for $\vect{k}=\vect{x},\vect{y},\vect{z}$. 
We can thus rewrite $\hat{J}_{\vect{n}}=\oplus_{N} \hat{J}_{\vect{n}}^{(N)}$,
where $\hat{J}_{\vect{n}}^{(N)} =\hat{\pi}_N \, \hat{J}_{\vect{n}} \, \hat{\pi}_N = 
\sum_{l=1}^N \hat{\sigma}_{\vect{n}}^{(l)}/2$ and
$\hat{\sigma}_{\vect{n}}^{(l)}$ is the Pauli matrix (along the direction $\vect{n}$
in the Bloch sphere, $\hat{\sigma}_{\vect{n}}^{(l)} = \alpha \hat{\sigma}_{\vect{x}}^{(l)} 
+ \beta \hat{\sigma}_{\vect{y}}^{(l)} + \gamma \hat{\sigma}_{\vect{z}}^{(l)}$) 
acting on the $l$-th particle.

The most general U(2) transformation, Eq.~(\ref{ModeOperator}), can be rewritten as 
\be \label{TMtransformU2}
\hat{\mathrm{U}}(\theta_1,\theta_2) = e^{i \chi \hat{J}_{\vect{s}}} 
\Big[ e^{-i \theta_1 \hat{a}_1^{\dag} \hat{a}_1} e^{-i \theta_2 \hat{a}_2^{\dag} \hat{a}_2} \Big] 
e^{-i \chi \hat{J}_{\vect{s}}},
\ee
where $\theta_1 = \phi_0+\theta/2$, $\theta_2 = \phi_0-\theta/2$,
$\vect{s}$ is a direction perpendicular to $\vect{z}$ and $\vect{n}$,
and $\cos \chi= \vect{n} \cdot \vect{z}$.
Equation~(\ref{TMtransformU2}) highlights the presence of two phases, 
$\theta_1$ and $\theta_2$, which can be identified as the phases acquired in each mode 
$a_1$ and $a_2$ inside a Mach-Zehnder-like interferometer 
[with standard balanced beam splitters replaced by the 
transformation $e^{\pm i \chi \hat{J}_{\vect{s}}}$, see Fig.~\ref{fig:Interf}(a)]. 
Both phases may be unknown.
When setting one of the two phases to zero (or to any fixed known value),
Eq.~(\ref{TMtransformU2}) reduces to different single-phase transformations:
\begin{itemize}

  \item SU(2) transformations $e^{-i \theta \hat J_{\vect{n}}}$ ($\phi_0=0$) or, equivalently
 \be \label{TMtransformSU2}
 \hat{\mathrm{U}}(\theta) = e^{+i \chi \hat{J}_{\vect{s}}} 
 e^{-i \theta \hat{J}_{\vect{z}}} e^{-i \chi \hat{J}_{\vect{s}}},
 \ee
 with notation analogous to Eq.~(\ref{TMtransformU2}) [see also Fig.~\ref{fig:Interf}(b)].
 This depends only on the relative phase shift $\theta=\theta_1-\theta_2$
 among the two interferometer modes.
 This encompasses the beam-splitter  
 $e^{-i\theta \hat{J}_{\vect{x}}}$, 
 the relative phase-shift  
 $e^{-i\theta \hat{J}_{\vect{z}}}$ 
 and the Mach-Zehnder 
 $e^{-i\theta \hat{J}_{\vect{y}}}$
 transformations.
 
 \item U(1) transformations $e^{-i \phi_0 \hat N}$ ($\theta=0$), which can be 
 understood as a phase shift equally imprinted on each of the two modes: 
 $e^{-i \phi_0 \hat N} = e^{-i \phi_0 \hat{a}^{\dag}_1 \hat{a}_1} \otimes e^{-i \phi_0 \hat{a}^{\dag}_2 \hat{a}_2}$.
 
\end{itemize}


\subsection{Output measurement}

Generally speaking, a POVM $\{\hat E(\varepsilon)\}_\varepsilon$
may or may not contain coherences among different number of particles.
A POVM does not contain number coherences if and only if all its elements $\hat E(\varepsilon)$
commute with the number of particles operator,
\be \label{Exicmomm}
[\hat E(\varepsilon),\hat N] = 0.
\ee
Equation~(\ref{Exicmomm}) is equivalent to~\cite{nota:commutatorE} 
\be \label{NM_POVM}
\hat E(\varepsilon)= \sum_N  \hat E_N(\varepsilon),
\ee
where $\hat E_N(\varepsilon) \equiv \hat \pi_N \, \hat E(\varepsilon) \, \hat \pi_N$
acts on the fixed-$N$ subspace and
$\hat \pi_N$ are projectors.

In current phase estimation experiments, the phase shift is estimated by measuring 
a function $f(N_1,N_2)$ of the number of particles at the output modes of the interferometer.
The \emph{experimentally relevant} POVMs can thus be written as 
\be
\hat E(\varepsilon) = \sum_{N_1,N_2} \delta\left[ f(N_1,N_2) - \varepsilon \right] \ket{N_1,N_2} \bra{N_1,N_2}.
\ee
By making a change of variable $N=N_1+N_2$ and $M=(N_1-N_2)/2$ ($-N/2\leq M \leq N/2$), 
we can rewrite this equation as 
\be
\hat E(\varepsilon) = \sum_N \sum_{M} \delta\left[ f(N,M) - \varepsilon \right] \ket{N,M} \bra{N,M},
\ee
which has the form of Eq.~(\ref{NM_POVM}).
Notice that the information about the total number of particles is not necessarily included in the POVM. 
For instance, the POVM corresponding to the measurement of only the relative number of particles can be written as
\be 
\hat E_M = \sum_N \proj{N,M},  \nonumber
\ee
which, again, has the form of Eq.~(\ref{NM_POVM}).
This example can be straightforwardly generalized to the measurement of any function of the 
relative number of particles. 
For the measurement of the number of particles in a single output port of the interferometer
(for instance at the output port ``1''), we have 
\be
\hat E_{N_{1}}  = \sum_{N,M} \delta\left[ N/2+M - N_1 \right] \, \proj{N,M}. \nonumber
\ee
We recover Eq.~(\ref{NM_POVM}) also in this case.
Analogous results hold for any function of $N_{1}$ (or $N_{2}$), for instance 
the measurement of the parity~\cite{GerryCONTPHYS2010} at one output port, $f(N_1,N_2)=(-1)^{N_1}$.


\subsection{Conditional probabilities}

For U(2) transformations, Eq.~(\ref{TMtransformU2}), 
the conditional probability can be written as
\be \label{U2CondProb}
P(\varepsilon \vert \theta_1, \theta_2) = \tr\left[ \hat E(\varepsilon) 
\hat{\mathrm{U}}(\theta_1,\theta_2)
\hat \rho \hat{\mathrm{U}}^\dag(\theta_1,\theta_2)\right].
\ee
If the probe state and/or the POVM do not contain number coherences, 
\ie~$\hat \rho$ is given by Eq.~(\ref{eq:incoherent}) and/or
$\hat E(\varepsilon)$ is given by Eq.~(\ref{NM_POVM}), 
then~(\ref{U2CondProb}) reduces to 
\be \label{Peps}
P(\varepsilon|\theta) = 
\sum_{N} Q_N P(\varepsilon \vert  N, \theta),
\ee
where $P(\varepsilon \vert  N, \theta) = 
\tr \big[\hat E(\varepsilon) e^{-i \theta \hat J_{\vect{n}}} 
\hat \rho^{(N)} e^{+i \theta \hat J_{\vect{n}}} \big]$.
The derivation of Eq.~(\ref{Peps}) is detailed in Appendix~B.
Equation~(\ref{Peps}) depends only on $\theta$, 
the relative phase shift among the two modes of the interferometer.
We conclude that U(2) transformations are relevant only 
if the input state contains coherences among different number of particles 
{\it and} the output measurement is a POVM with coherences.
In all other cases the phase shift $e^{-i\phi_0 \hat N}$ is irrelevant 
as the conditional probabilities are insensitive to $\phi_0$. 
In this case, the mode transformation Eq.~(\ref{TMtransformU2}) 
restricts to the unimodular ({\it i.e.} unit determinant)
subgroup SU(2).
The SU(2) representation, while being not general, is widely used
because, in current experiments, the phase shift is 
estimated by measuring a function of the number of particles at 
the output ports of the interferometer. 
Table \ref{tableU} summarizes the general two-mode transformation 
group for the phase estimation problem, depending on the presence 
of number coherences in the probe state and POVM.

\begin{table}[h!] 
\begin{center}
\begin{tabular}{|c|c|c|}
\hline     
     & POVM with coh.            & POVM without coh.  \\
\hline
$\hat \rho$ with coh.     & U(2)                 & SU(2) \\
\hline
$\hat \rho$ without coh.  & SU(2)                & SU(2) \\
\hline
\end{tabular}
\caption{The table summarizes the general two-mode transformation group for the phase estimation problem. 
The U(2) group is only relevant when number coherences are present in both the probe state {\it and} in the POVM.} \label{tableU}
\end{center}
\end{table}


\subsection{Multiphase estimation}
\label{Sec.Estimation}

Since U(2) transformations involve two phases, $\theta_1$ and $\theta_2$,  
we review here the theory of two-parameter estimation \cite{PezzeVarenna}.
The vector parameter $\vect{\theta} \equiv \{ \theta_1, \theta_2\}$
is inferred from the values $\vect{\varepsilon} \equiv \{ \varepsilon_1, \varepsilon_2, ..., \varepsilon_m\}$
obtained in $m$ repeated independent measurements. 
The mapping from the measurement results into the two-dimensional parameter space is 
provided by the estimator function 
$\vect{\Theta}(\vect{\varepsilon}) \equiv [\Theta_1(\vect{\varepsilon}),\Theta_2(\vect{\varepsilon})]$.
Its mean value is $\bar{\vect{\Theta}} \equiv [ \bar{\Theta}_1, \bar{\Theta}_2]$, 
with $\bar{\Theta}_i = \int \ud \vect{\varepsilon} 
\, \mathcal{L}(\vect{\varepsilon}|\vect{\theta}) 
\Theta_i(\vect{\varepsilon})$ ($i=1,2$) and
the likelihood function $\mathcal{L}(\vect{\varepsilon}|\theta) \equiv \prod_{l=1}^m P(\varepsilon_l|\vect{\theta})$.
We further introduce the covariance matrix $\mathbf{B}$ of elements
\be
\mathbf{B}_{i,j} = \int \ud \vect{\varepsilon} 
\, \mathcal{L}(\vect{\varepsilon}|\vect{\theta}) 
\big(\Theta_i(\vect{\varepsilon}) - \bar{\Theta}_i \big)
\big(\Theta_j(\vect{\varepsilon}) - \bar{\Theta}_j \big).
\ee 
Notice that $\mathbf{B}$ is symmetric and its $i$th
diagonal element is the variance $(\Delta \Theta_i)^2$ of $\Theta_i(\varepsilon)$.

\subsubsection{Cram\'er-Rao bound}

Following a Cauchy-Schwarz inequality~\cite{Kay_book}, we have~\cite{note:saturation}:
\be \label{CSCR}
\big(\vect{v}^{\top} \mathbf{b} \, \vect{u} \big)^2 \leq m 
\big(\vect{v}^{\top} \mathbf{B} \vect{v} \big) \big(\vect{u}^{\top} \mathbf{F} \vect{u} \big), 
\quad \forall \, \vect{u}, \vect{v} \in \ensuremath{\mathbbm R},
\ee
where $\mathbf{b}_{i,j} = \partial \bar{\Theta}_i /\partial \theta_j$ is the Jacobian matrix 
and 
\be \label{componentsFmatrix}
\mathbf{F}_{i,j} = \int \ud \vect{\varepsilon} \, \frac{1}{P(\vect{\varepsilon}\vert \vect{\theta})}
\bigg( \frac{\partial P(\vect{\varepsilon}\vert \vect{\theta})}{\partial \theta_i} \bigg) 
\bigg( \frac{\partial P(\vect{\varepsilon}\vert \vect{\theta})}{\partial \theta_j} \bigg)
\ee
the Fisher information matrix~\cite{notaFij}, which is symmetric and nonnegative definite.
Note that $\mathbf{B}$, $\mathbf{F}$ and $\mathbf{b}$ generally depend on $\vect{\theta}$ 
but we do not explicitly indicate this dependence, in order to simplify the notation.
Note also that $\mathbf{b}$ may depend on $m$.
In the inequality~(\ref{CSCR}) $\vect{u}$ and $\vect{v}$ are  \emph{arbitrary} real vectors.
Depending on $\vect{v}$ and $\vect{u}$ we thus have an infinite number of scalar inequalities.
If the Fisher matrix is positive definite, and thus invertible,  
the specific choice $\vect{u} = \mathbf{F}^{-1} \mathbf{b}^{\top} \vect{v}$ 
in Eq.~(\ref{CSCR}) leads to the vector parameter Cram\'er-Rao lower bound 
$\mathbf{B} \geq \mathbf{B}_{\rm CR}$~\cite{CramerBOOK},
in the sense that the matrix $\mathbf{B} - \mathbf{B}_{\rm CR}$ is 
nonnegative definite [i.e. $\vect{v}^{\top} \mathbf{B} \vect{v} \geq \vect{v}^{\top} \mathbf{B}_{\rm CR} \vect{v}$
holds for all real vectors $\vect{v}$], where 
\be \label{multiCR}
\mathbf{B}_{\rm CR} = \frac{\mathbf{b} \, \mathbf{F}^{-1} \, \mathbf{b}^{\top}}{m}.
\ee 
This specific choice of $\vect{u}$ 
leads to a bound which is saturable by the maximum likelihood estimator (see Sec.~\ref{Sec:ML})
asymptotically in the number of measurements.
 
In the two-parameter case, the Fisher information matrix 
\be
\mathbf{F} = \left[
\begin{array}{cc}
 F_{1,1} & F_{1,2} \\
 F_{1,2} & F_{2,2} \\ 
\end{array} 
\right]
\ee
is invertible if and only if $F_{1,1}F_{2,2} - F_{1,2}^2 \neq 0$, its inverse given by 
\be
\mathbf{F}^{-1} = \frac{1}{F_{1,1}F_{2,2} - F_{1,2}^2} 
\left[
\begin{array}{cc}
 F_{2,2} & -F_{1,2} \\
 -F_{1,2} & F_{1,1} \\ 
\end{array} 
\right].
\ee
Furthermore, if $\bar{\vect{\Theta}}_i$
does not depend on $\theta_j$ for $j \neq i$
({\it i.e.} $\mathbf{b}$ is diagonal), 
the diagonal elements of $\mathbf{B}_{\rm CR}$ satisfy the inequalities:
\beq \label{ineqF}
(\Delta \theta_i)^2_{\rm CR} &=&
\frac{F_{j,j} \mathbf{b}_{i,i}^2 }{m(F_{i,i}F_{j,j} - F_{i,j}^2)} \geq 
\frac{\mathbf{b}_{i,i}^2}{mF_{i,i}}, \label{Fii} 
\eeq
with $i\neq j$, $i,j=1,2$.
For the two-parameter case, the inequality~(\ref{ineqF}) can be immediately demonstrated 
by using $F_{1,1}F_{2,2} - F_{1,2}^2 > 0$ which holds since $\mathbf{F}$ nonnegative definite and 
assumed here to be invertible.

In the estimation of a single parameter, the matrix $\mathbf{B}_{\rm CR}$
reduces to the variance $(\Delta \theta_{\rm CR})^2$. 
Equation~(\ref{multiCR}) becomes
\be \label{CR}
(\Delta \theta_{\rm CR})^2 = \frac{b^2}{m F},
\ee
where $b \equiv \ud \bar{\Theta}(\theta)/\ud \theta$
[for unbiased estimators $b =1$, {\it i.e.} $\bar{\Theta}(\theta) = \theta$] and 
$F =\int d\varepsilon \, \frac{1}{P(\varepsilon|\theta)}
\big( \frac{\ud P(\varepsilon|\theta)}{\ud \theta}\big)^2$ is the (scalar) Ficher information (FI).
By comparing Eq.~(\ref{Fii}) and 
Eq.~(\ref{CR}), we see, as reasonably expected, that the estimation uncertainty 
of a multi-parameter problem is always larger or at most equal than the uncertainty obtained for a  
single parameter (namely, when all other parameters are exactly known).


\subsubsection{Maximum likelihood estimation}
\label{Sec:ML}

A main goal of parameter estimation is to find the estimators saturating the Cram\'er-Rao bound.
These are called efficient estimators.
While such estimators are rare, it is not  possible to exclude, in general, that an efficient unbiased 
estimator may exist for any value of $m$. 
One of the most important estimators is the 
maximum likelihood (ML)
$\vect{\Theta}_{\rm ML}( \vect{\varepsilon})$.
It is defined as the value $\vect{\Theta}_{\rm ML}( \vect{\varepsilon})$ which maximizes the log-likelihood 
function:
\be
\vect{\Theta}_{\rm ML}( \vect{\varepsilon}) =
\arg\Big[ \max_{\vect{\varphi}} \log \mathcal{L}(\vect{\varepsilon}|\vect{\varphi}) \Big].
\ee
It is possible to demonstrate, by using the law of large numbers and the central 
limit theorem, that, asymptotically in the number of measurements, the maximum likelihood is unbiased
and normally distributed with variance given by the inverse Fisher information matrix~\cite{PezzeVarenna, Kay_book}. 
Therefore, the specific choice of vector $\vect{u}$ which leads to the 
Cram\'er-Rao bound~(\ref{multiCR}) is justified by the fact that the ML saturates this bound for a 
sufficiently large number of measurements. 
  
    
\subsubsection{Quantum Cram\'er-Rao bound}

The Fisher information matrix, satisfies
\be \label{FQMineq}
\mathbf{F} \leq \mathbf{F}_{\rm Q},
\ee
in the sense that the matrix $\mathbf{F}_{\rm Q} - \mathbf{F}$ is
positive definite.
The symmetric matrix $\mathbf{F}_Q$ is called the quantum Fisher information matrix and 
its elements are
\be \label{componentsFQmatrix}
[\mathbf{F}_{\rm Q}]_{i,j} = \frac{1}{2} \tr\Big[\hat \rho(\theta) \big(\hat L_{i}\hat L_{j}+\hat L_{j} \hat L_{i}\big) \Big],
\ee
with $i=1,2$, where the self-adjoint operator $\hat L_{i}$, called 
the symmetric logarithmic derivative (SLD) \cite{Helstrom_book}, is defined as
\be \label{SLD1}
\frac{\partial \hat \rho(\vect{\theta})}{\partial \theta_i} = \frac{\hat L_{i} \hat \rho(\vect{\theta}) 
+ \hat \rho(\vect{\theta}) \hat L_{i}}{2}.
\ee
In particular, we have 
$\frac{\ud P(\varepsilon|\vect{\theta})}{\ud \theta_i} 
= \Re(\tr[\rho(\vect{\theta}) \hat{E}(\varepsilon) \hat L_{i}])$,
$\Re(x)$ being the real part of $x$.
Note also that the operator $\hat L_{i}$ (and also $\mathbf{F}_{\rm Q}$) generally depends on $\vect{\theta}$.
Equation (\ref{FQMineq}) holds for any Fisher information matrix (invertible or not)
and there is no guarantee that, in general, the equality sign can be saturated. 
Assuming that $\mathbf{F}$ and $\mathbf{F}_{\rm Q}$ are positive definite (and thus invertible)
and combining Eq.~(\ref{multiCR}) -- in the unbiased case -- with Eq.~(\ref{FQMineq}), 
we obtain the matrix inequality $\mathbf{B}_{\rm CR} \geq \mathbf{B}_{\rm QCR}$ \cite{nota_invposmat}, where 
\be \label{multiQCR}
\mathbf{B}_{\rm QCR} = \frac{\mathbf{F_{\rm Q}}^{-1}}{m}.
\ee
This sets a fundamental bound, the quantum Cram\'er-Rao (QCR) bound~\cite{Helstrom_book}, 
for the sensitivity of unbiased estimators.
The bound cannot be saturated, in general, in the multiparameter case.

In the single parameter case, we have 
$\Delta \theta_{\rm CR} \geq \Delta \theta_{\rm QCR}$, where 
\be \label{QCR}
\Delta \theta_{\rm QCR} = \frac{1}{\sqrt{m F_Q[\hat\rho(\theta)]}}.
\ee
The (scalar) quantum Fisher information (QFI) can be written as 
\be \label{QFIdefL}
F_Q[\hat\rho(\theta)]=(\Delta \hat L)^2,
\ee 
where $\hat L$ is the $\theta$-dependent SLD and 
we used $\tr[\hat \rho(\theta) \hat L] = 0$.
The equality $\Delta \theta_{\rm CR} = \Delta \theta_{\rm QCR}$ (or, equivalently $F=F_Q$) 
holds if the POVM $\{ \hat E(\varepsilon) \}$
is made by the set of projector operators over  
the eigenvectors of the operator $\hat L$, 
as first discussed in Ref.~\cite{Braunstein_1994}.
The quantum Cram\'er-Rao is a very convenient way to calculate the phase uncertainty since
it only depends on the properties of the probe state and not on the quantum measurement.
  
  
\section{Fisher Information for states without number coherences}
\label{BoundsFisherSU2}
  
As discussed above, for states without number coherences we can restrict to SU(2) transformations 
and thus the estimation of a single parameter: the relative phase shift among the arms of a Mach-Zehnder-like 
interferometer.
In this case, an important property of the QFI holds:
\be \label{FisherNoCoh}
F_\mathrm{Q} \big[\hat\rho_\mathrm{inc}, \hat{J}_{\vect{n}}\big] 
= \sum_N Q_N F_\mathrm{Q} \big[\hat{\rho}^{(N)}, \hat{J}_{\vect{n}}^{(N)}\big],
\ee 
where $F_Q[\hat \rho^{(N)}, \hat J_{\vect{n}}^{(N)}]$ is the QFI calculated on 
the fixed-$N$ subspace.
To demonstrate this equation
let us consider the general expression of QFI given in Ref.~\cite{Braunstein_1994}, 
\be 
F_Q[\hat \rho, \hat J_{\vect{n}}]= 2 
\sum_{\substack{ i,j \\ p_i+p_j \neq 0 } }
\frac{(p_i-p_j)^2}{p_i+p_j} 
\big\vert \langle i \vert \hat J_{\vect{n}} \vert j \rangle \big\vert^2,
\ee
where $p_j \geq 0$ and $\{ \ket{j} \}$ is a basis of the Hilbert space, 
$\sum_j \vert j \rangle \langle j \vert = \Eins$, chosen such that
$\hat \rho = \sum_j p_j \vert j \rangle \langle j \vert$. 
For states without number coherence, we have 
$\hat \rho_{\rm inc} = \sum_N Q_N \sum_j p_j^{(N)} \vert j^{(N)} \rangle \langle j^{(N)} \vert$
where $\{\vert j^{(N)} \rangle\}$ is a basis on the fixed-$N$ subspace.
Since $\langle j^{(N)} \vert \hat J_{\vect{n}} \vert j'^{(N')} \rangle = 
\langle j^{(N)} \vert \hat J^{(N)}_{\vect{n}} \vert j'^{(N)} \rangle \delta_{N,N'}$, 
{\it i.e.} $\hat J_{\vect{n}}$ does not couple states of different number of particles. 
In an analogous way it is possible to demonstrate that the SLD
$\hat{L} = \sum_N \hat{L}^{(N)}$.
We thus conclude that, when the input state does not have number coherences, 
the Von Neumann measurement on the eigenstates of $\hat L^{(N)}$ for each value of $N$ 
-- which in particular does not have number coherences -- is such that the corresponding FI 
saturates the QFI.


\section{Fisher Information for states with number coherences}
\label{BoundsFisherU2}

In this section we discuss the quantum Fisher information 
for states with number coherences. 
First we consider the estimation a single phase, either $\phi_0$ or $\theta$,
separately, assuming that the other parameter is known.
We then apply the multiparameter estimation theory outlined above to 
calculate the sensitivity when $\theta_1$ and $\theta_2$ are both estimated at the same time.
We will mainly focus on the calculation of an upper bound to the quantum Fisher information.

\subsection{Single parameter estimation}

Let us consider the different transformations outlined
in Sec.~\ref{Sec:TwoMode}:

$\bullet$ SU(2) transformation $\hat U = e^{-i \theta \hat J_{\vect{n}} }$.
It is interesting to point out that, for SU(2) transformations, 
number coherences may increase the value of the QFI.
We have 
\be \label{FishIneqCoh}
F_Q\big[ \ket{\psi}, \hat J_{\vect{n}} \big] \geq F_Q\big[ \hat{\rho}_{\rm inc}, \hat J_{\vect{n}} \big], \nonumber
\ee
where $\ket{\psi}=\sum_N \sqrt{Q_N} \, \ket{\psi_N}$ is a normalized pure state with coherences
and $\hat{\rho}_{\rm inc} = \sum_N \hat \pi_N \proj{\psi} \hat \pi_N=
\sum_N \vert Q_N \vert \,  \proj{\psi_N}$
is obtained from $\proj{\psi}$ by tracing out the number-coherences.
Notice that, if $F_Q[ \ket{\psi}, \hat J_{\vect{n}} ] > F_Q [ \hat{\rho}_{\rm inc}, \hat J_{\vect{n}} ]$ holds,  
then that saturation of $F_Q[ \ket{\psi}, \hat J_{\vect{n}} ]$
necessarily requires a POVM with number coherences. 
This is a consequence of the fact that the Fisher information obtained 
with POVMs without coherences is independent on the presence of 
number coherences in the probe state and it is therefore 
upper bounded by $F_Q [ \hat{\rho}_{\rm inc}, \hat J_{\vect{n}} ]$.
Equation (\ref{FishIneqCoh}) can be demonstrated using {\it i})
$F_Q[ \ket{\psi}, \hat J_{\vect{n}} ] = 4 (\Delta \hat J_{\vect{n}} )^2_{\ket{\psi}}$ \cite{Braunstein_1994, Pezze_2009}
and 
$F_Q\big[ \hat{\rho}_{\rm inc}, \hat J_{\vect{n}} \big] 
= \sum_N Q_N (\Delta \hat{J}^{(N)}_{\vect{n}} )_{\ket{\psi_N}}^2$ [see Eq.~(\ref{FisherNoCoh})], 
where we have explicitly indicated the state on which the variance is calculated on 
(we will keep this notation where necessary and drop it elsewhere)
and {\it ii}) the Cauchy-Schwartz inequality 
\be \label{CSINEQ}
  \bigg(\sum_N \vert Q_N \vert \bra{\psi_N}  \hat J_{\vect{n}}^{(N)} \ket{\psi_N}\bigg)^2
  \le \sum_{N} \vert Q_{N} \vert \bra{\psi_N} \hat J_{\vect{n}}^{(N)} \ket{\psi_N}^2. 
\ee
The equality holds if and only if $\bra{\psi_N} \hat J_{\vect{n}}^{(N)} \ket{\psi_N}$ 
is a constant independent of $N$. 

In the following we discuss the bounds to the QFI.
For this, a useful property of the QFI is its convexity \cite{PezzeVarenna}.
In our case it implies 
\be \label{FQineq}
F_Q[\hat\rho_{\rm coh}, \hat J_{\vect{n}} ] \leq 
\sum_k p_k F_Q\big[\ket{\psi_k}, \hat J_{\vect{n}} \big] = 
4 \sum_k p_k  
(\Delta \hat J_{\vect{n}} )^2_{\ket{\psi_k}},
\ee
where the equality holds only for pure states. 
Furthermore,
\beq \label{DeltaJn}
4(\Delta \hat J_{\vect{n}})^2_{\ket{\psi_k}} 
&\leq& 
4 \sum_N \vert Q_{N,k} \vert \langle J_{\vect{n}}^2 \rangle_{\ket{\psi_{N,k}}} \nonumber \\
&\leq& 
\sum_N \vert Q_{N,k} \vert N^2 = \langle \hat{N}^2 \rangle_{\ket{\psi_k}}. 
\eeq
The first inequality is saturated for $\langle \hat J_{\vect{n}} \rangle_{\ket{\psi_{k}}} = 0$.
In the second inequality we used 
$4\langle \hat J_{\vect{n}}^2 \rangle_{\ket{\psi_{N,k}}} \leq N^2$
both saturated for the NOON state
$\vert {\rm NOON}_{ \vect{n} \rangle } \equiv (\vert N,0 \rangle_{\vect{n}} + \vert 0,N \rangle_{\vect{n}})/\sqrt{2}$
with $\hat J_{\vect{n}} \vert N,0 \rangle_{\vect{n}} = (N/2) \vert N,0 \rangle_{\vect{n}}$
[and $\hat J_{\vect{n}} \vert 0,N \rangle_{\vect{n}} = -(N/2) \vert 0,N \rangle_{\vect{n}}$].
In this case, by using Eq.~(\ref{FQineq}), we have that 
\be
F_Q[\hat\rho_{\rm coh}, \hat{J}_{\vect{n}}] \leq \tr\big[ \hat \rho_{\rm coh} \hat N^2 \big]
\ee 
where the equality can be saturated by a coherent superposition of NOON states
(note indeed that $\bra{{\rm NOON}_{\vect{n}}} \hat J_{\vect{n}} \ket{{\rm NOON}_{\vect{n}}} = 0$).
We thus have 
\be
(\Delta \theta)^2_{\rm QCR} \geq \frac{1}{m \tr\big[\hat \rho_{\rm coh} \hat N^2\big]}.
\ee

$\bullet$ U(2) transformations $\hat U = e^{-i \phi_0 \hat N}$. 
Using the convexity of the QFI, we have 
\beq 
F_Q[\hat\rho_{\rm coh}, \hat{N}] &\leq& 4\sum_k p_k 
\big( \Delta \hat N \big)^2_{\ket{\psi_k}} 
\leq 4 \big( \Delta \hat N \big)_{\hat\rho_{\rm coh}}^2 
\label{FineqCoh}
\eeq
where the second inequality follows from a Cauchy-Schwarz inequality.
We thus have 
\be
(\Delta \phi_0)^2_{\rm QCR} \geq \frac{1}{4 m \big( \Delta \hat N \big)_{\hat\rho_{\rm coh}}^2}.
\ee

\subsection{Two-parameter estimation}

In the U(2) framework, there are, in general, two phases to estimate: $\phi_0$ and $\theta$. 
When estimating both at the same time, the phase sensitivity is calculated using 
the multiphase estimation formalism discussed above. 
The inequality (\ref{ineqF}), leads to
\be
(\Delta \theta)^2_{\rm CR} \geq \frac{1}{m F_Q[\rho_{\rm coh}, \hat{J}_{\vect n} ]}, 
\quad 
(\Delta \phi_0)^2_{\rm CR} > \frac{1}{m F_Q[\rho_{\rm coh}, \hat{N} ]}, \nonumber
\ee
which can be further bounded by using the above inequalities for the QFI. 
For pure states we have  
\begin{equation*}
\mathbf{F}_Q^{-1} = 
\frac{2}{\mathrm{det}[\mathbf{F_Q}]}
\left(
\begin{array}{cc}
2(\Delta \hat{J}_{\vect{n}} )^2 & \mean{\hat N} \mean{ \hat{J}_{\vect{n}}} - \mean{\hat N \hat{J}_{\vect{n}}}  \\
\mean{\hat N} \mean{ \hat{J}_{\vect{n}}} - \mean{\hat N \hat{J}_{\vect{n}}} & (\Delta \hat N)^2/2
\end{array} \right), \nonumber
\end{equation*}
where $\mathrm{det}[\mathbf{F_Q}] = 4(\Delta \hat N)^2(\Delta \hat{J}_{\vect{n}} )^2 - 4[\mean{\hat N \hat{J}_{\vect{n}}} - \mean{\hat N} \mean{ \hat{J}_{\vect{n}} }]^2$.
We thus have 
\be \label{ineqN}
(\Delta \phi_0)^2 \geq \frac{m^{-1}}{(\Delta \hat N)^2 - [\mean{\hat N \hat{J}_{\vect{n}}} - \mean{\hat N} \mean{ \hat{J}_{\vect{n}} }]^2/(\Delta \hat{J}_{\vect{n}} )^2}, 
\ee
which, in particular, is always larger than $1/m(\Delta \hat N)^2$, and 
\be \label{ineqJn}
(\Delta \theta)^2 \geq \frac{m^{-1}}{4(\Delta \hat{J}_{\vect{n}} )^2 - 4[\mean{\hat N \hat{J}_{\vect{n}}} - \mean{\hat N} \mean{ \hat{J}_{\vect{n}} }]^2/(\Delta \hat N)^2}, 
\ee
which is always larger than $1/4m(\Delta \hat{J}_{\vect{n}} )^2$.


\section{Separability and Entanglement}
\label{Ent1} 

When the number of particles is fixed, there exists a precise relation between 
the entanglement properties of a probe state and the QFI:
if the state is separable [{\it i.e.} can be written as in Eq.~(\ref{rhosep})]
then the inequality 
\be
F_Q\big[\hat\rho^{(N)}_{\rm sep},{\hat J}_{\vect{n}}^{(N)}\big] \leq N
\ee
holds \cite{Pezze_2009}.
A QFI larger than $N$ is a sufficient condition for entanglement and singles out 
the states which are useful for quantum interferometry, {\it i.e.} states that 
can be used to achieve a sub shot noise phase uncertainty.
The above inequality can be extended to the case of multiparticle entanglement.
In Refs~\cite{HyllusArXiv10,TothArXiv10}, it has been shown that 
for $k$-producible states the bound
\be
	\label{eq:FQ_class}
	F_Q[ \rho_{k-{\rm prod}};{\hat J}_{\vect{n}}^{(N)} ]\le s k^2 + r^2
\ee
holds, where $s=\lfloor \frac{N}{k}\rfloor$ 
is the largest integer smaller than
or equal to $\frac{N}{k}$ and $r=N-sk$.
Hence a violation of the bound~(\ref{eq:FQ_class}) proves
$(k+1)$-particle entanglement.
For general states of a fixed number of particles, we have 
$F_Q[\hat\rho^{(N)},{\hat J}_{\vect{n}}^{(N)}] \leq N^2$ \cite{Pezze_2009, Giovannetti_2006}, 
whose saturation requires $N$-particle entanglement.

In the case of states with number fluctuations, 
the situation is more involved.
For states without number coherences, by using Eq.~(\ref{FisherNoCoh}), 
we straightforwardly obtain 
\be \label{QFIseparable}
F_Q\big[\hat{\rho}_{\rm sep}, \hat{J}_{\vect{n}}\big] = \sum_N Q_N F_Q\big[\hat\rho^{(N)}_{\rm sep},
{\hat J}_{\vect{n}}^{(N)}\big]
\leq \sum_N Q_N N = \langle N \rangle. 
\ee
The phase sensitivity achievable with separable states without number coherences thus 
satisfies the chain of inequalities 
$\Delta \theta \leq \Delta \theta_{\rm CR} \leq \Delta \theta_{\rm QCR} \leq \Delta \theta_{\mathrm{SN}}$,
where 
\begin{equation} \label{SN}
\Delta \theta_{\mathrm{SN}} = \frac{1}{\sqrt{ m\langle N \rangle }},
\end{equation}
which agrees with the common definition of the shot-noise or
standard quantum limit.
This brings us to the following results. An arbitrary state with non-fixed number 
of particles {\em but without number-coherences} 
is entangled if it fulfils the inequality 
\be \label{chi2}
	\chi^2 \equiv \frac{\langle \hat N \rangle}{F_Q[\hat{\rho},\hat{J}_{\vect{n}}]} <1,
\ee
for some direction $\vect{n}$. 
Entanglement is a necessary resource for sub shot-noise sensitivity in linear SU(2) interferometers,
{\it i.e.} when number coherences are not available or not measured.
States $\hat{\rho}$ satisfying Eq.~(\ref{chi2}) 
are useful in a linear interferometer implemented by the transformation 
$\hat{J}_{\vect{n}}$, since, according to Eq.~(\ref{QCR}), 
they can provide a sub shot-noise (SSN) phase sensitivity.

The relation between the properties of a probe state without number coherences and 
the QFI can be further extended to the case of multiparticle entanglement.
Using Eqs.~(\ref{FisherNoCoh}) and (\ref{rho_inc}), we have 
$F_Q[\rho_{k-{\rm prod}}^{\rm inc};\hat J_{\vect{n}}]=\sum_N Q_N F_Q[\rho_{k-{\rm prod}}^{(N)};\hat J_{\vect{n}}^{(N)}]$ 
and thus, by using Eq.~(\ref{eq:FQ_class}),
\bean \label{QFImultipart}
	F_Q[\rho^{\rm inc}_{k-{\rm prod}};\hat J_{\vect{n}}]
	&\le& \sum_N Q_N \left( \lfloor \frac{N}{k}\rfloor k^2+(N-\lfloor \frac{N}{k}\rfloor k)^2\right)\\
	&=& \mean{\hat s k^2}+\mean{\hat r^2}.
\eean
Here, $\hat s=\lfloor \frac{\hat N}{k}\rfloor$, 
$\hat r=\hat N-\hat sk$, 
commute with the number operator $\hat N$. 
The maximum value of the Fisher information
is thus obtained for maximally entangled states ($k=N$) and is  
\be \label{GeneralFisherNoCoh}
\max_{\hat{\rho_{\rm inc}}} F_Q[\hat{\rho}_{\rm inc}, \hat{J}_{\vect{n}}] = \langle N^2 \rangle.
\ee
Equation (\ref{GeneralFisherNoCoh}) is reached 
for incoherent superpositions of NOON states
$\sum Q_N \vert \rm{NOON} \rangle_{\vect{n}}\langle \rm{NOON} \vert$.
By using Eq.~(\ref{GeneralFisherNoCoh}), we can define an upper limit 
for the quantum Cram\'er-Rao bound Eq.~(\ref{QCR}), maximized over all possible quantum states:
\be \label{maxQCR}
\min_{\hat{\rho_{\rm inc}}}  \Delta \theta_{\rm{QCR}} = \frac{1}{\sqrt{m \langle N^2 \rangle}}.
\ee
In particular it is always true that $\langle N^2 \rangle \geq \langle N \rangle^2$, where 
the equality holds if and only if the number of particles does not fluctuate.
By a proper choice of the $Q_N$ distribution, $\langle N^2 \rangle$
can be an arbitrary function of $\langle N \rangle$.
Therefore, when fixing $\langle N \rangle$, the bound Eq.~(\ref{maxQCR})
can be arbitrarily small, even zero for distribution having $\langle N^2 \rangle = + \infty$.
This was first noticed in Ref.~\cite{Hofmann_2009}. 
The significance of the bound Eq.~(\ref{maxQCR}) is the subject of a vivid debate in the 
recent literature~\cite{AnisimovPRL10, ZhangJPA2012}.

Let us now turn to the case of states with number coherences. 
In this case there is no clear relation between separability/multiparticle-entanglement 
(as defined in Sec.~\ref{SubSecEnt}) and the QFI.
The two examples below illustrate this fact: we show states with 
number coherences which are separable in each subspace of finite number of particles
and have a QFI that can be arbitrarily larger than $\langle N \rangle$.
In other words, the inequality $\Delta \theta_{\rm QCR} \leq \Delta \theta_{\mathrm{SN}}$ 
does not hold for separable states with number coherences.
However, it is still possible to find a bound of phase sensitivity 
for SU(2) transformations if we restrict to POVMs without number coherences. 
In this (non-optimal) case, the Fisher information calculated for $k$-producible states, Eq.~(\ref{rho_inc}), is 
$F \leq \mean{\hat s k^2}+\mean{\hat r^2}$. 
In particular, the Fisher information calculated for separable states, Eq.~(\ref{sep}), is 
$F \leq \mean{\hat{N}}$. 
Therefore, for POVM without coherences and coherent separable states,   
the chain of inequalities $\Delta \theta \leq \Delta \theta_{\rm CR} \leq \Delta \theta_{\mathrm{SN}}$
holds. 
States with number coherences which, in a phase estimation 
experiment using POVMs without number coherences, overcome the shot noise sensitivity, 
are necessary entangled within the definition given in Sec.~\ref{SubSecEnt}.
 

\subsection{Examples}

\subsubsection{Example: MOON state} 
\label{Example:1}

Let us consider the state (which we call ``MOON'' state) 
\be \label{MOON}
\vert \psi \rangle = \sqrt{\frac{N}{N+M}} \, e^{i\phi} \, \vert M,0\rangle + 
\sqrt{\frac{M}{N+M}} \,\vert 0,N\rangle,
\ee 
with $N,M>0$.
For $N \neq M$ this state is separable in each subspace of a fixed number of particles
and thus separable according to our definition~(\ref{sep}).
If $N=M$ Eq.~(\ref{MOON}) reduces to the well known NOON state, which is maximally entangled.
The QFI is maximum along the $\vect{z}$ direction and given by 
\be \label{QFIexaple1}
F_Q\big[\vert \psi \rangle,\hat J_z\big] = N \, M. 
\ee
The average number of particles in (\ref{MOON}) is
$\langle \hat N \rangle = 2NM/(N + M)$.
Therefore, since $N+M>2$, we have $F_Q\big[\vert \psi \rangle,\hat J_z\big] > \langle \hat N \rangle $. 
We thus have an example of a separable state (with number coherences)
which has a QFI larger than the average number of particles.
The incoherent mixture is obtained from the pure state Eq.~(\ref{MOON}) by projecting over fixed-$N$
subspaces is
\be
\hat{\rho} = \frac{N}{N+M} \proj{M,0} + \frac{M}{N+M} \proj{0,N}.
\ee
Its QFI which is maximum on the plane orthogonal to $\vect{z}$
and fulfils $F_Q\big[\hat \rho,\hat J_{\vect{n}} \big] \leq \langle \hat N \rangle$,
as expected.
Notice that
$F_Q\big[\vert \psi \rangle,\hat J_z\big] \geq F_Q\big[\hat \rho,\hat J_z\big]$, as expected.


\subsubsection{Example: coherence with the vacuum} 
\label{Example:2}

An example similar to the one above has been discussed by
Benatti and Braun in Ref. \cite{BenattiPRA2013} and highlights how the coherence
with the vacuum state can increase the QFI (see also \cite{Rivas_2011, PezzePRA2013}, in the single-mode case). 
Let us take
\be \label{ExampleBB}
\vert \psi \rangle = \sqrt{1 - \frac{\langle \hat N \rangle}{N}} \, \vert 0,0\rangle + 
\sqrt{\frac{\langle \hat N \rangle}{N}} \, e^{i\phi_N}\, \vert N,0\rangle,
\ee
where $\langle \hat N \rangle \leq N$ is the average number of particles.
The QFI for rotations around the $\hat J_z$ axis is 
$F_Q\big[\vert \psi \rangle, \hat J_z\big] 
= N\langle \hat N \rangle - \langle \hat N \rangle^2$.
By properly choosing $N$ it is possible to reach arbitrary large values of the QFI.
For instance $F_Q\big[\vert \psi \rangle, \hat J_z\big] > \langle \hat N \rangle^k$
for $N > \langle \hat N \rangle^{k-1}+\langle \hat N \rangle$ and any $k>0$.
In particular, Eq.~(\ref{ExampleBB}) is, as above, an example of separable state [according to Eq. (\ref{sep})]
which has a QFI larger than $\langle \hat N \rangle$ [for $N > \langle \hat N \rangle +1$].
We also have $F_Q\big[\vert \psi \rangle, \hat J_z\big] > \langle \hat N \rangle^2$
for $N > 2 \langle \hat N \rangle$. 
Finally note that, as expected, the condition $F_Q\big[\vert \psi \rangle, \hat J_z\big] < \langle N^2 \rangle = N \langle N \rangle$ is always fulfilled (for $\langle N \rangle>0$).


\section{The Heisenberg limit}
\label{SecHL}

\begin{figure}[t!]
\begin{center}
\includegraphics[scale=0.5]{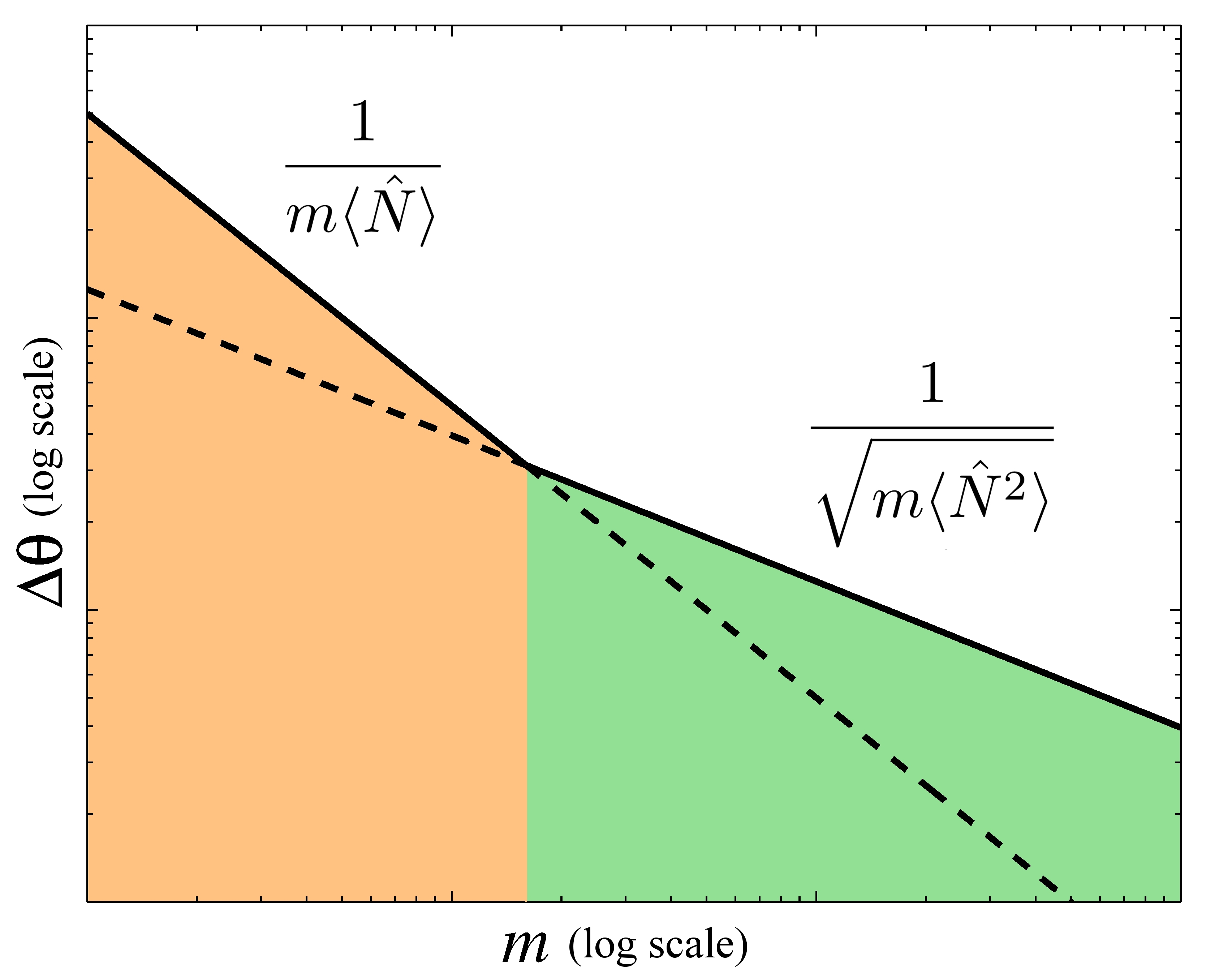}
\end{center}
\caption{ (color online) Schematic representation of Eq.~(\ref{HL}) in loglog scale (solid line). 
In the left hand side of the figure (orange region),
$\Delta \theta_{\rm HL} = 1/m \langle \hat N \rangle$ (solid line), which is is larger than 
$1/\sqrt{m \langle \hat N^2 \rangle}$ (dashed line). 
In the right hand side of the figure (green region), for $m \geq \langle N^2 \rangle/\langle N \rangle^2$, 
$\Delta \theta_{\rm HL} 1/\sqrt{m \langle \hat N^2 \rangle}$ (solid line) is larger than $1/m \langle \hat N \rangle$ (dashed line)
}  \label{fig:HL}
\end{figure}

In this section we discuss the ultimate phase sensitivity allowed when fixing the average number of particles 
$\langle \hat{N} \rangle$ in the probe state.
This is generally indicated as the Heisenberg limit.
We focus on SU(2) transformations~$e^{-i \theta \hat J_{\vect{n}}}$.
We show that the Heisenberg limit for states and/or POVMs without number coherences is given by
Eq.~(\ref{HL}).
The first bound in Eq.~(\ref{HL})~will be demonstrated in the following. It is generally not tight and is valid 
for estimators which are unbiased in each fixed-$N$ subspace [see Sec.~\ref{HL_POVM_nocoh}].
The second bound in Eq.~(\ref{HL}) is the optimal quantum Cram\'er-Rao bound, Eq.~(\ref{maxQCR}), 
for estimators which are globally unbiased ($\bar \Theta = \theta$).
It can be saturated by the maximum likelihood estimator in the central limit ({\it i.e.} for $m\gtrsim m_{\rm{cl}}$ 
and a sufficiently large $m_{\rm{cl}}$) by using an incoherent mixture of NOON states. 
Since $\langle \hat N^2 \rangle \geq \langle \hat N \rangle$, the first bound in Eq.~(\ref{HL}) 
is significant for small values of $m$. 
Comparing the two bounds in Eq.~(\ref{HL}) one obtains a lower bound for $m_{\rm{cl}}$,
\be \label{mcl}
m_{\rm{cl}} \geq \frac{\langle N^2 \rangle}{\langle N \rangle^2}. 
\ee
Therefore, the larger is $\langle N^2 \rangle$, the smaller is the 
Eq.~(\ref{maxQCR}) but, at the same time, the larger is the number of repeated measurements needed
to reach the central limit and saturate Eq.~(\ref{maxQCR}).
If $\langle N^2 \rangle  \to \infty$, reaching the central limit 
requires an infinite number of measurement $m \to +\infty$, and, accordingly, 
the phase uncertainty vanishes, $\Delta \theta \to 0$.
Equation (\ref{HL}) is schematically represented in Fig.~\ref{fig:HL}.
For states and POVM with number coherences we give some argument on the validity of 
Eq.~(\ref{HL}), even though a conclusive demonstration is missing.
For this case, we give an overview of the results obtained in the literature.

\subsection{Heisenberg limit for states and/or POVMs without number coherences}
\label{HL_POVM_nocoh}

The mean value (for $m$ independent measurements) of an arbitrary estimator $\Theta(\vect{\varepsilon})$ is
\beq \label{Eq:HLdim1}
\bar\Theta &=& 
\int \ud \vect{\varepsilon} \, 
P(\vect{\varepsilon}|\theta) \, \Theta(\vect{\varepsilon}), 
\eeq
where $\vect{\varepsilon}\equiv\{\varepsilon_1,\varepsilon_2,...,\varepsilon_m\}$ and
$P(\vect{\varepsilon}|\theta)=\prod_{i=1}^m P(\varepsilon_i|\theta) $.
Taking $P(\varepsilon|\theta)$ as in Eq.~(\ref{Peps}),
we can rewrite Eq.~(\ref{Eq:HLdim1}) as 
\be \label{Thetaest1}
\bar\Theta=\sum_{\vect{N}} \, Q_{\vect{N}} \, \bar{\Theta}_{\vect{N}},
\ee
where the sum extends over all possible sequences
$\vect{N}\equiv\{N_1,N_2,...,N_m\}$, 
$Q_{\vect{N}} \equiv \prod_{i=1}^{m} Q_{N_i}$ 
is the probability of the given sequence [$\sum_{\vect{N}} Q_{\vect{N}}=1$],
\be
\bar{\Theta}_{\vect{N}} \equiv 
\int \ud \vect{\varepsilon} \,
P( \vect{\varepsilon} | \vect{N},\theta) \, 
\Theta( \vect{\varepsilon}),
\ee
and $P(\vect{\varepsilon}| \vect{N},\theta) \equiv \prod_{i=1}^{m} P(\varepsilon_i| N_i,\theta)$.
Following an analogous method,  
we can rewrite the standard deviation of the estimator as
\beq \label{Dthest}
	(\Delta \theta)^2
	&=&\sum_{\vect{\varepsilon}}P(\vect{\varepsilon}|\theta) \, 
	\left[ \Theta(\vect{\varepsilon}) - \bar\Theta\right]^2 \nonumber \\
	&=& \sum_{\vect{N}} Q_{\vect{N}} \left[ \bar{\Theta}_{\vect{N}} - \bar{\Theta} \right]^2
	+ \sum_{\vect{N}} Q_{\vect{N}} \big(\Delta\Theta_{\vect{N}}\big)^2, \,\,\,\,\,\,\,\,\,\,\,
\eeq
where 
\be
\big(\Delta\Theta_{\vect{N}}\big)^2 \equiv 
\int \ud \vect{\varepsilon} \,
P( \vect{\varepsilon} | \vect{N},\theta) \, 
\left[ \Theta( \vect{\varepsilon}) - \bar\Theta_{\vect{N}}\right]^2
\ee
is the variance of the estimator $\Theta( \vect{\varepsilon})$ for a given 
sequence $\vect{N}$.
Since $\int \ud \vect{\varepsilon} P(\vect{\varepsilon}| \vect{N},\theta)=1$,
we can apply the Cram\'er-Rao theorem to set a bound to the variance 
$\big(\Delta\Theta_{\vect{N}}\big)^2$:
\be \label{CRN}
\big(\Delta\Theta_{\vect{N}}\big)^2 \geq 
\frac{b_{\vect{N}}^2}{F_{\vect{N}} (\theta)} ,
\ee
where $b_{\vect{N}} \equiv \partial_\theta \bar\Theta_{\vect{N}}$
and
\be
F_{\vect{N}} (\theta) \equiv \int \ud \vect{\varepsilon} \, \frac{1}{ P(\vect{\varepsilon}| \vect{N},\theta) } 
\left( \frac{\ud}{\ud \theta} P(\vect{\varepsilon}| \vect{N},\theta) \right)^2
\ee
is the Fisher information for the specific sequence $\vect{N}$.
Note that $F_{\vect{N}} (\theta) = \sum_{i=1}^m F_{N_i}(\theta)$, where 
$F_{N_i}(\theta)$ is the Fisher information calculated on the subspace of
$N_i$ particles, 
\be
F_{N_i}(\theta) = 
\int \ud \varepsilon_i \,
\frac{1}{P(\varepsilon_i \vert N_i,\theta)} 
\left( \frac{\ud}{\ud \theta} P(\varepsilon_i \vert N_i,\theta) \right)^2.
\ee
If all the numbers $N_i$ are equal to $N$, we would recover 
$F_{\vect{N}} (\theta) = m F_{N}(\theta)$ and thus the usual multiplication factor $m$. 
Note also that the Fisher information $F_{N_i}(\theta)$ is bounded as
$F_{N_i}(\theta) \leq N_i^2$ \cite{Giovannetti_2006, Pezze_2009}, 
and thus $F_{\vect{N}} (\theta) \leq \sum_{i=1}^m N_i^2=\vect{N} \cdot \vect{N}=\vect{N}^2$.
By using this result and Eqs.~(\ref{CRN}) we obtain
\be \label{Dthineq1}
\sum_{\vect{N}} Q_{\vect{N}} \big(\Delta\Theta_{\vect{N}}\big)^2 \geq 
\sum_{\vect{N}} \frac{Q_{\vect{N}} b_{\vect{N}}^2}{\vect{N}^2}
\geq 
\sum_{\vect{N}} \frac{Q_{\vect{N}} b_{\vect{N}}^2}{\mathcal{S}(\vect{N})^2}
\ee
where $\mathcal{S}(\vect{N}) \equiv \sum_{i=1}^m N_i$ is the sum of 
all values of $N$ in the sequence,  and 
we have used the inequality $\vect{N}^2 \leq \mathcal{S}(\vect{N})^2$ which follows since the $N_i$ are all positive numbers.
We now use the Cauchy-Schwarz inequality 
\be
\sum_{\vect{N}} \frac{ Q_{\vect{N}} b_{\vect{N}}^2 }{\mathcal{S}(\vect{N})^2} \, 
\sum_{\vect{N}'} Q_{\vect{N}'} \geq 
\left( \sum_{\vect{N}} \frac{Q_{\vect{N}} \, b_{\vect{N}}}{ \mathcal{S}(\vect{N}) } \right)^2.
\ee
Using the normalization of $Q_{\vect{N}}$ and Eq.~(\ref{Dthineq1}), we have 
\be \label{Dthineq2}
\sum_{\vect{N}} Q_{\vect{N}} \big(\Delta\theta_{\vect{N}}\big)^2 \geq 
\left( \sum_{\vect{N}} \frac{Q_{\vect{N}} \, b_{\vect{N}}}{ \mathcal{S}(\vect{N}) } \right)^2.
\ee
A second Cauchy-Schwarz inequality,
\be
\sum_{\vect{N}} \frac{Q_{\vect{N}} \, b_{\vect{N}}}{ \mathcal{S}(\vect{N}) } \, 
\sum_{\vect{N}'} Q_{\vect{N}'} \mathcal{S}(\vect{N}') \geq
\left( \sum_{\vect{N}} Q_{\vect{N}} \sqrt{b_{\vect{N}}}\right)^2,
\ee
where we note that $\sum_{\vect{N}} Q_{\vect{N}} \mathcal{S}(\vect{N})= m \langle \hat N \rangle$ and 
$b_{\vect{N}}$ are positive numbers, gives
\be \label{Dthineq3}
\sum_{\vect{N}} Q_{\vect{N}} (\Delta\theta_{\vect{N}})^2 \geq 
\frac{\Big( \sum_{\vect{N}} Q_{\vect{N}} \, \sqrt{b_{\vect{N}}} \, \Big)^4}{\big(m \langle \hat N \rangle\big)^2}.
\ee
Finally, by using Eqs.~(\ref{Dthest}) and (\ref{Dthineq3}), the sensitivity 
of the estimator can be bounded by
\be \label{Dthineqfinal}
(\Delta \theta)^2 \geq 
\sum_{\vect{N}} Q_{\vect{N}} \left( \bar{\Theta}_{\vect{N}} - \bar{\Theta} \right)^2
+ \frac{\Big( \sum_{\vect{N}} Q_{\vect{N}} \, \sqrt{b_{\vect{N}}} \, \Big)^4}
{\big(m \langle \hat N \rangle\big)^2}.
\ee
This is the main result of this section.
The first term in Eq.~(\ref{Dthineqfinal}) is always positive and is characteristic of  
phase estimation with probe states of a non-fixed number of particles. 
It is equal to zero
if and only if $\bar \Theta_{\vect{N}} = \bar{\Theta}$ for all possible sequences $\vect{N}$. 
Since sequences with $m$ values of the same total number of particles $N$ are possible, 
the above condition 
implies that the mean value of the estimator is 
the same (and equal to $\bar{\Theta}$) in each fixed-$N$ subspace.
Furthermore, a convenient situation is to have an unbiased 
estimator $\bar{\Theta} = \theta$ for all values of $m$. In this case 
(assuming that the first term in Eq.~(\ref{Dthineqfinal}) is 
equal to zero, $\bar{\Theta}_{\vect{N}} =\theta$ for all possible sequences $\vect{N}$), 
we have \cite{Hyllus_2010}
\be \label{Dthineqfinalfinal}
\Delta \theta \geq 
\frac{1}{m \langle \hat N \rangle}.
\ee
We recall that this holds when the 
estimator is unbiased in each fixed-$N$ subspace and for all the values of $m$.
If this does not hold, the more general, but less conclusive, inequality~(\ref{Dthineqfinal})
can be used.


\subsection{Some considerations about the Heisenberg limit 
for states and POVMs with number coherences}
\label{HLCL}

In this subsection we show that the bound $1/m\mean{\hat N}$
applies in the fully coherent situation \emph{at least} in the central limit.

As discussed in Sec.~\ref{BoundsFisherU2}, 
the optimal quantum Cram\'er-Rao bound is
$\Delta \theta_{\rm QCR} = 1/\sqrt{m \tr[\hat \rho_{\rm coh} \hat N^2] }$, 
which is uniquely saturated by 
a probe given by superpositions of pure NOON states of the form 
$\vert \psi \rangle = \sum_N \sqrt{Q_N} \, \ket{ {\rm NOON }_{\vect{n}} }$ [see discussion after Eq.~(\ref{DeltaJn})].
Since $\bra{{\rm NOON}_{\vect{n}}} \hat J_{\vect{n}} \ket{{\rm NOON}_{\vect{n}}} = 0$ for any $N$,
the QFI can be written, in this case, as $F_Q = 4 \mean { \hat J_{\vect{n}}^2 }$.
In addition, since the operator $\hat J_{\vect{n}}$ commutes with $\hat N$, 
off-diagonal terms $N \neq N'$ in the density matrix $\ket{{\rm NOON}_{\vect{n}}}\bra{{\rm NOON}_{\vect{n}}}$
do not play any role in the calculation of the Cram\'er-Rao bound.
Hence a mixture 
$\sum_N Q_N \ketbra{{\rm NOON}_{\vect{n}}}$ reaches the same value of the quantum Fisher information. 
It follows that the limit $\Delta \Theta=1/\sqrt{m_{\rm cl} \mean{N^2}}$ can be saturated 
for $m\ge m_{\rm cl}^{\rm inc}$ with a POVM without coherences between states with different numbers of particles.
It may happen, however, that the number of measurements for which 
the Cram\'er-Rao bound is saturated is different if number coherences in the 
state and POVM are used.
For an asymptotically large $m$, the saturation in both cases 
is guaranteed by the Fisher theorem.
In this regime, the results of Sec.~\ref{HL_POVM_nocoh} hold and 
we conclude that the Heisenberg limit Eq.~(\ref{HL}) is valid in also in the full coherent case.
In particular, $\Delta\Theta\ge 1/m \langle \hat{N} \rangle$ is a general bound for
sufficiently large $m$, even if states with coherences
and POVMs with coherences are used.

A final remark concerns the uniqueness of the states saturating
the Cram\'er-Rao bound. Saturating $F_Q=\mean{\hat N^2}$ 
requires that $4(\Delta \hat J_{\vect{n}}^{(N)})^2=N^2$ for any $N$.
Only states such that $\hat\pi_N\rho\hat\pi_N=Q_N \vert\rm{NOON}\rangle_{\vect{n}}\langle \rm{NOON} \vert$
satisfy this constraint. However, there are many such states
for given $\mean{\hat N}$ and $\mean{\hat N^2}$.
This is because fixing $\mean{\hat N}$ and $\mean{\hat N^2}$ corresponds to 
choosing two constraints on the distribution $\{Q_N\}$, in 
addition to the constraints $\sum_N Q_N =1$ and $Q_N\ge 0$. 
$\{Q_N\}$ is in general not uniquely defined by these
constraints.
Even though, even though different state can have the same QFI  
$F_Q=\mean{\hat N^2}$, they may be characterized by different 
values of $m_{\rm cl}$, {\it i.e.}  the minimal number 
where the central limit is reached, 
depending also on the estimator. 


\subsection{Overview of the recent literature \\ on the Heisenberg limit}

The comparison between our results and the recent literature deserves some discussion. 
We recall that our definition of Heisenberg limit, Eq.~(\ref{HL}), 
holds for two-mode transformations and unbiased estimators.
It does not apply to states and POVM with number coherences outside the central limit, 
for which no conclusive results has been obtained so far. 
A summary of our findings is reported in Table \ref{TableHL}.
In the literature, the problem of defining the Heisenberg limit for states and POVMs with number coherences 
has been tackled with different techniques which we briefly discuss below. 
Overall, there is a general strong indication that Eq.~(\ref{HL}) is the general form of Heisenberg limit.
While the literature leaves open the possibility to overcome the bound $1/m \langle \hat N \rangle$
at specific phase values (called ``sweet spots'', see below),
there is no proposal showing convincing evidences of sub-Heisenberg uncertainties. 

\begin{table}[b!]
\begin{center}
\begin{tabular}{|c|c|c|}
\hline
 & POVM with coh. & POVM without coh. \\
\hline
$\hat \rho_{\rm in}$ with coh.   & $\Delta \theta \geq \Delta \theta_{\rm CR}$ & 
$\Delta \theta \geq \Delta \theta_{\rm HL}$ \\
\hline
$\hat \rho_{\rm in}$ without coh.  & 
$\Delta \theta \geq \Delta \theta_{\rm HL}$ & 
$\Delta \theta \geq \Delta \theta_{\rm HL}$ \\
\hline
\end{tabular}
\caption{Table summarizing the fundamental bounds of phase sensitivity
discussed in this manuscript.
For states and/or POVM without number coherence, the Heisenberg 
limit is given by the competition of two
bounds [see Eq.~(\ref{HL})], as explained in Sec.~\ref{SecHL}. 
For general POVMs and states with number-coherences ({\it i.e.} for U(2) transformations) 
only the Cram\`er-Rao bound applies.
In this case and for SU(2) transformations, the Heisenberg limit Eq.~(\ref{HL})
holds at least in the central limit, as discussed in Sec.~\ref{HLCL}.} \label{TableHL}
\end{center}
\end{table}

Before presenting an overview of the literature, it is important to recall here that there 
are two models of phase estimation: {\it i})~The first model assumes that
the phase to be estimate is a nonrandom unknown quantity. 
This is the framework discussed in this manuscript. 
It assumes that we can collect an arbitrary number of sequences 
$\vect{\varepsilon} = \{ \varepsilon_1, \varepsilon_2, ..., \varepsilon_m \}$ of 
$m$ measurements while keeping fixed the (unknown) phase shift in the apparatus. 
The phase sensitivity is given by the variance of the estimator $\Theta(\vect{\varepsilon})$
(see Sec.~\ref{Sec.Estimation}):
\be \label{NoBaysens1}
(\Delta \Theta)^2_\theta = \int \ud \vect{\varepsilon} \, P(\vect{\varepsilon} \vert \theta)  \big[ \Theta(\vect{\varepsilon}) - \bar{\Theta}(\theta) \big]^2,
\ee
where $\bar{\Theta}(\theta) $ is the $\theta$-dependent mean value of the estimator. 
{\it ii})~The second model assumes that the phase is a random variable with a probability
distribution $P(\theta)$ called ``the prior''. 
Parameter estimation based on this model is referred to as Bayesian estimation \cite{VanTreesBOOK}.
In this case, each sequence $\vect{\varepsilon}$ of $m$ measurements is obtained with a phase shifts 
randomly varying with a probability $P(\theta)$. 
The phase sensitivity is defined as the weighed mean square error
\be \label{Baysens1}
(\Delta \Theta)^2_{\rm bay} =  \int \ud \theta \int \ud \vect{\varepsilon} \, P(\vect{\varepsilon} , \theta) \big[ \Theta(\vect{\varepsilon}) - \bar{\Theta}(\theta) \big]^2,  
\ee 
where $P(\vect{\varepsilon} , \theta)=P(\vect{\varepsilon} \vert \theta) P(\theta)$ is the joint probability distribution of phase $\theta$ and experimental measurement $\vect{\varepsilon}$.

\subsubsection{Sweet spot phase estimation} 

Let us consider here the estimation of a fixed phase shift
with a state and POVM with number coherences.
In this case the Cram\`er-Rao is the sole sensitivity bound. 
At certain phase values  (indicated as ``sweet spots'' in Ref.~\cite{GLM_preprint2011}) 
it can be arbitrary small when fixing the average number of particles 
$\langle \hat{N} \rangle$ in the state.
Nevertheless, in Ref.~\cite{GLM_preprint2011} it is shown that 
the sum of sensitivities calculated at two nearby phase shifts, $\theta_1$ and $\theta_2$,
is bounded when the phases are sufficiently far apart. 
For unbiased estimators, the inequality \cite{GLM_preprint2011} 
\be \label{NoBaysens2}
\frac{ (\Delta \Theta)_{\theta_1}+ (\Delta \Theta)_{\theta_2} }{2} \geq 
\frac{\kappa}{m(\langle \hat{H} \rangle - H_0)}
\ee
holds, where $\hat{H}$ is the generator of phase shift 
(the phase encoding transformation $e^{-i \hat H \theta}$ is assumed), 
$H_0$ is the minimum eigenvalue of $\hat{H}$ populated in the probe state. 
The maximum value of $\kappa$ is 0.074 reached when $\vert \theta_1 - \theta_2 \vert \geq 0.83/m(\langle \hat{H} \rangle - H_0)$.
In the special (which yet might be non optimal) case when the phase sensitivity $(\Delta \Theta)_{\theta}$
does not depend on $\theta$, Eq.~(\ref{NoBaysens2}) 
implies $(\Delta \Theta)_{\theta} \geq \kappa/m(\langle \hat{H} \rangle - H_0)$ \cite{GLM_preprint2011}.
These results require the generator of phase shift to have a discrete spectrum and a finite lowest eigenvalue \cite{GLM_preprint2011}.  
The bound (\ref{NoBaysens2}) thus holds, for instance, 
for single-mode phase estimation, when the generator of phase shift is the number of particles 
operator, $\hat{N}$. 
Equation~(\ref{NoBaysens2}) does not hold for the two mode case unless one imposes a 
bound on the total number of particles distribution.   
It should also be noticed that the bound found in Ref.~\cite{GLM_preprint2011} refers to the mean square 
fluctuation of the estimator with respect to the true phase values. 
It coincides to Eq.~(\ref{NoBaysens1}) only if the estimator is unbiased. 
For biased estimators, the bound (\ref{NoBaysens2}) does not hold.

\subsubsection{Bayesian bounds}

Several works~\cite{Tsang_preprint2011, Giovannetti_2012, HallPRA2012, HallNJP2012, BerryPRA2012}
have discussed the Heisenberg limit within the framework Bayesian phase estimation, {\it i.e.}
when the phase sensitivity is averaged over the prior, Eq.~(\ref{Baysens1}).
This approach might be considered as a generalisation of the averaging over two phases discussed above \cite{GLM_preprint2011}. 
In this case, the Heisenberg limit is found by making use of suitable Bayesian bounds.  
Using the Ziv-Zakai (Bayesian) bound~\cite{VanTreesBOOK} in the ``low prior information regime'' 
(e.g. when $P(\phi)$ is uniform a phase interval sufficiently wider than $1/m\langle \hat{H} \rangle$)
it was possible to demonstrate that~\cite{Tsang_preprint2011, Giovannetti_2012}
\be \label{dphiZZ}
(\Delta \Theta)_{\rm bay}  \geq \frac{\alpha}{m\mean{\hat{H}}},
\ee 
where $\alpha$ is a constant~\cite{Tsang_preprint2011, Giovannetti_2012}
($\alpha=0.1548$ for an uniform prior distribution~\cite{GaoJPA2012}). 
In the opposite regime, when the width of $P(\phi)$ is smaller than $1/m\langle \hat{H} \rangle$,
the phase uncertainty is essentially determined by the prior distribution \cite{Giovannetti_2012, GaoJPA2012}.  
In this case, sub-Heisenberg uncertainties are possible but ineffective
({\it i.e.} the estimation process does not bring more information than a random guess of the phase 
within the prior $P(\theta)$ itself \cite{Giovannetti_2012}).  
In Refs.~\cite{Tsang_preprint2011, Giovannetti_2012} the bound~(\ref{dphiZZ}) 
was demonstrated by assuming $\hat{H}$ to have a finite lower bound in the spectrum, as  
in the single-mode case with $\hat{H}=\hat{N}$.
The extension of Eq.~(\ref{dphiZZ}) to unbounded Hamiltonians (and thus when $\hat H = \hat J_{\vect{n}}$)
is discussed in \cite{GaoJPA2012}.

In Refs.~\cite{HallPRA2012, HallNJP2012, BerryPRA2012}, 
using an entropic uncertainty relation, 
it was possible to show that
\be \label{dphiER}
(\Delta \Theta)_{\rm bay}  \geq \frac{\beta}{m\langle{ |\hat{H}-h}| \rangle}, 
\ee
where $h$ is an arbitrary eigenvalue of $\hat{H}$, which can have a discrete or continuous spectrum \cite{HallNJP2012}, 
and $\beta$, depending on the prior distribution $P(\phi)$, can be arbitrarily small for 
a sufficiently narrow prior ($\beta = 0.559$ for a completely random phase shift in a $2\pi$ interval \cite{HallPRA2012}).  
The derivation of Eq.~(\ref{dphiER}) does not require $\hat{H}$ to be discrete, have integer eigenvalues 
or have a lowest eigenvalue \cite{HallNJP2012}.
In particular, the bound applies for two-mode operators \cite{BerryPRA2012}, {\it i.e} when $\hat{H} = \hat J_{\vect{n}}$. 
In this case, we have
$\langle \psi \vert \, \vert \hat{J}_{\vect{n}} \vert \, \vert \psi \rangle 
= \sum_N \sum_{\mu=-N/2}^{N/2} \vert \mu \vert \, \vert Q_{N, \mu} \vert^2$, 
where $-N/2 \leq \mu \leq N/2$ are eigenvalues of $\hat{J}_{\vect{n}}^{(N)}$ with eigenstate $\vert N, \mu \rangle$
and $\vert \psi \rangle = \sum_{N,\mu} Q_{N,\mu} \vert N, \mu \rangle$ is a state with number coherences. 
Using $\vert \mu \vert \leq N/2$ we obtain  
$\langle \vert \hat{J}_{\vect{n}} \vert \rangle \leq \langle \hat N \rangle /2$ and thus, from Eq.~(\ref{dphiER}), 
$\delta \phi  \geq \beta/m\langle \hat N \rangle $.

\subsubsection{Proposal by Rivas and Luis} 

Reference~\cite{Rivas_2011} discusses a single-mode phase estimation 
reaching, at specific phase values, a phase uncertainty arbitrarily 
smaller than $1/m\langle \hat{N} \rangle$.
This claim is the result of a calculation of the Fisher information for states
with strong coherences with the vacuum (see also the example in Sec.~\ref{Example:2}). 
Rivas and Luis argue that the maximum likelihood estimator might reach 
an arbitrary small phase uncertainty \cite{Rivas_2011}. 
Results and claims similar to the one of Ref.~\cite{Rivas_2011} can be found in the early
literature \cite{Shapiro_1989, Shapiro_1991,Schleich_1990}.
The bounds (\ref{dphiZZ}) and (\ref{dphiER}) do not apply to this case and therefore there are no analytical results in the 
literature that forbid the conclusions of Ref.~\cite{Rivas_2011} (and also of \cite{Shapiro_1989, Shapiro_1991,Schleich_1990}). 
A detailed numerical analysis of the estimation protocol proposed in Ref.~\cite{Rivas_2011} can be found in \cite{PezzePRA2013},
showing no violation of the Heisenberg limit.
It is shown \cite{PezzePRA2013} that the number of measurements needed to saturate an arbitrary small Cram\'er-Rao bound  
is so large that the Heisenberg limit $\Delta \theta = 1/m\langle\hat{N}\rangle$ is not overcome.
Analogous conclusions were reported in Ref.~\cite{Braunstein_1992},
showing no violation of the Heisenberg limit for the proposals \cite{Shapiro_1989, Shapiro_1991,Schleich_1990}. 


\subsection{Examples}

\subsubsection{Example: biased estimator} 
We recall once again that the demonstration of Eq.~(\ref{HL}) reported in Sec.~\ref{HL_POVM_nocoh} 
requires the estimator to be unbiased in each fixed-$N$ subspace. 
If this is not the case, the bound $1/m \langle \hat{N} \rangle$ can be violated. 
This is explicitly shown in the following example.
We consider the state
\be
\hat \rho = (1-p) \, \vert 0,0 \rangle \langle 0,0 \vert + p \, \vert \psi_M \rangle \langle \psi_M \vert,
\ee
where $\vert \psi_M \rangle$ is, for instance, a NOON state of $M$ particles, 
$\vert \psi_M \rangle = \big(\vert M,0\rangle + \vert 0,M \rangle\big)/\sqrt{2}$.
The average number of particles is
$\langle \hat N \rangle = p M$. 
The QFI is 
\be
F_Q[\hat \rho, \hat J_{\vect{n}}] = p F_Q\Big[\vert \psi_M \rangle, \hat J_{\vect{n}}\Big] = pM^2 =
\frac{\langle \hat N \rangle^2}{p},
\ee
leading to the quantum Cram\'er-Rao bound 
\be
\big(\Delta \theta_{QCR}\big)^2 = \frac{1}{m F_Q[\hat \rho, \hat J_{\vect{n}}]} =
\frac{p}{m \langle \hat N \rangle^2}.
\ee
We assume to have an estimator such that \cite{Rafal_Example}
\begin{equation*}
\Theta(\varepsilon) = \left\{
\begin{array}{ll}
0 & \text{  if $N=0$} \\
\tilde \Theta(\varepsilon)/p & \text{  if $N=M$}. \\
\end{array} \right.
\end{equation*}
where $\varepsilon$ is the result of a possible measurement in the fixed-$N$ subspace
and $\tilde \Theta(\varepsilon)$ is an arbitrary unbiased estimator.
The estimator is biased on each $N$ subspace but it is globally unbiased,
$\bar{\Theta} = \theta$.
Let us consider a single measurements ($m=1$), the standard deviation of the estimator is given by Eq.~(\ref{Dthest})
\beq
(\Delta \Theta)^2&=&
	\sum_{N} Q_{N} \Big( \bar{\Theta}_{N} - \bar{\Theta} \Big)^2
	+ \sum_{N} Q_{N} \big(\Delta\Theta_{N}\big)^2 = \nonumber \\
	&=& \bigg( \frac{1-p}{p}\bigg) \theta^2 + \frac{(\Delta \tilde \Theta)^2}{p} \nonumber\\
	&\geq& \bigg( \frac{1-p}{p}\bigg) \theta^2 + \frac{1}{pF_Q\big[\vert \psi_M \rangle, \hat J_{\vec n}\big]} \nonumber \\
	&\geq&  \bigg( \frac{1-p}{p}\bigg) \theta^2 + \frac{p}{\langle \hat N \rangle^2}.
\eeq
The first term highlights the role of $\theta=0$ as a sweet spot for the phase estimation.
If $\theta=0$ we may have a violation of Eq.~(\ref{HL}) -- by an arbitrary small factor $p$ --
due to the fact that the estimator is biased in each fixed-$N$ subspace.

\subsubsection{Example: Two-mode SSW state} 
\label{ExampleSSW}

In 1989 Shapiro, Shephard and Wong \cite{Shapiro_1989, Shapiro_1991, Schleich_1990} 
proposed a state (hereafter indicated as SSW state) 
that can be used to overcome the limit $\Delta \theta = 1/\mean{\hat N}$ in a single-mode phase estimation.
It was then showed by Braunstein et al. \cite{Braunstein_1992},
with a maximum likelihood analysis, that the SSW state does not allow to overcome 
$\Delta \theta =1/m\mean{\hat N} = 1/\bar{N}_{tot}$ in the central limit.
The analysis of \cite{Braunstein_1992} emphasizes the non-trivial role of 
the number of repeated measurements, $m$.
We here extend the SSW state to two modes and study how our bounds apply to this case.

The SSW state can be straightforwardly extended to two-mode 
as, for instance, a superposition of Twin-Fock states:  
\be \label{TMSSW}
\ket{\psi_{\rm SSW}} = \sum_{n=0}^M \frac{A}{n+1} \ket{n,n},
\ee
where $A$ is a normalization constant, $M$ is a cut-off and
$\ket{n,n}$ is a Twin-Fock state \cite{Holland_1993} of $2n$ particles.
Using the results of Refs. \cite{Schleich_1990,Braunstein_1992}, we have
\beq
A^2 &=& \frac{6}{\pi^2} + \frac{36}{\pi^4(M+1)} + O\Big(\frac{1}{M^2}\Big), \nonumber \\
\Big\langle \frac{\hat N}{2}+1 \Big\rangle &=& \frac{6}{\pi^2} \big[ \gamma + \ln(M+1) \big] + O\Big(\frac{\ln M}{M}\Big), 
\nonumber \\
\Big\langle \Big( \frac{\hat N}{2}+1 \Big)^2 \Big\rangle &=& A^2 (M+1), \nonumber
\eeq
where $\gamma \approx 0.57721$ is the Euler's constant.
To the leading order in $M$, the QFI is given by
\beq
F_Q\left[\ket{\psi_{\rm SSW}},\hat J_y\right]  \approx  
\frac{12}{\pi^2} (M+1) 
\approx
\frac{12}{\pi^2} e^{-\gamma} e^{\frac{\pi^2}{6} \mean{\frac{\hat N}{2} +1}} \nonumber
\eeq
The state Eq.~(\ref{TMSSW}), similarly to its one-mode counterpart \cite{Braunstein_1992}, 
has a QFI that scales exponentially with $\mean{\hat N}$, diverging for $M \to \infty$.
The Quantum Cram\'er-Rao bound can thus be arbitrarily small.
What is the optimal sensitivity achievable with the SSW state, Eq.~(\ref{TMSSW}) ?
According to our results, if the phase is estimated with a POVM without coherences, 
the Heisenberg limit is given by Eq.~(\ref{HL}), which becomes, 
\be \label{SSWHL}
\Delta \theta_{\rm HL} = \max\bigg[ \frac{\pi e^{\frac{\gamma}{2}}}{2\sqrt{3m}} e^{-\frac{\pi^2}{12} \mean{\frac{\hat N}{2} +1}},\frac{1}{m \mean{\hat N}}\bigg]
\ee
and thus
\be \label{SSWmcl}
m_{\rm cl} \geq \frac{\pi^2 e^{-\gamma}}{12}  \frac{e^{\frac{\pi^2}{6} \mean{\frac{\hat N}{2} +1}}}{\mean{\hat N}^2}.
\ee
We can obtain an arbitrary high sensitivity but i) it is always larger than 
$1/m \mean{\hat N}$ (second term in Eq.~(\ref{SSWHL})) and, according to Eq.~(\ref{SSWmcl}), 
ii) the central limit is reached for a number of measurements 
$m_{\rm cl}$ which 
diverges for $\mean{\hat N} \to \infty$. This conclusion holds for any arbitrary unbiased estimator, 
even though the maximum likelihood remains the most relevant example.
If the phase is estimated by a POVM with coherences, Eq.~(\ref{SSWHL})
holds in the central limit, at least.

\subsubsection{Example: two-mode squeezed vacuum state} 

In Ref.~\cite{AnisimovPRL10} it was argued that the two-mode squeezed vacuum state
can be used to overcome the Heisenberg limit in a Mach-Zehnder interferometer with parity detection
in a single-output.
This example is similar to the one discussed above 
(In Sec.~\ref{ExampleSSW}) nevertheless it is worth 
analysing it in details. 
The two-mode squeezed vacuum state is:
\be \label{state:tmsv}
\ket{\psi} = \sum_{N=0}^{+\infty} \frac{ e^{-i \psi N} (\tanh r)^{N}}{\cosh r} \ket{N,N},
\ee 
where $r$ is a squeezing parameter. 
In optics, it can be experimentally produced by a nondegenerate down-conversion process
with a nonlinear crystal.
With atoms, a state similar to Eq.~(\ref{state:tmsv}) can be obtained with 
spin-dependent collisions in spinor Bose-Einstein condensates.
For the state Eq.~(\ref{state:tmsv}), we have $\mean{\hat N} = 2 \sinh^2 r$, 
$\mean{\hat N^2} = 2\mean{\hat N}\big(\mean{\hat N}+1\big)$
and 
$(\Delta N)^2 = \sinh^2 2r = \mean{\hat N}\big(\mean{\hat N}+2\big)$.
For a POVM without number coherences, as the one considered in~\cite{AnisimovPRL10},  
the Heisenberg limit~(\ref{HL}) is
\be \label{HLtmsq}
(\Delta \theta)_{\mathrm{HL}} = 
\max\bigg[ \frac{1}{\sqrt{2 m \mean{\hat N}\big(\mean{\hat N}+1\big)}}, \frac{1}{m \langle \hat N \rangle} \bigg].
\ee
This can be compared to the quantum Cram\'er-Rao bound for rotations around the $\vect{y}$ axis, 
[corresponding to the Mach-Zehnder interferometer transformation, with quantum Fisher information $F_Q = 4(\Delta \hat{J}_y)^2$]:
\be \label{CRtmsq}
(\Delta \theta)_{\mathrm{QCR}}
= \frac{1}{\sqrt{m \mean{\hat N}\big(\mean{\hat N}+2\big)}}.
\ee
In Ref.~\cite{AnisimovPRL10} the sensitivity was calculated with an error propagation 
formula and, at $\theta=0$, it matches Eq.~(\ref{CRtmsq}). 
Equation (\ref{CRtmsq}) overcomes $(\Delta \theta) = 1/\sqrt{m}\langle \hat N \rangle$
that is often indicated as the Heisenberg limit \cite{AnisimovPRL10}.
While this appears as the natural extension of Eq.~(\ref{HL_fixed}) 
to the case of fluctuating number of particles (by replacing $N$ with $\langle \hat{N} \rangle$), 
it is not a fundamental bound. 

In the large-$m$ limit Equation~(\ref{CRtmsq}) is always (for the interesting case $\mean{\hat N} > 1$) 
smaller than the Heisenberg limit Eq.~(\ref{HLtmsq}).
The two-mode squeezed vacuum state is very useful to overcome the shot noise limit
but it does not really surpass the Heisenberg limit (even if POVMs with number coherences are used). 
The saturation of the Heisenberg limit in the large $m$ limit Eq.~(\ref{HLtmsq}) can be 
obtained with the superposition of NOON states 
$\sum_{N=0}^{+\infty} \frac{ e^{-i \psi N} (\tanh r)^{N}}{\cosh r} \frac{\ket{N,0}+\ket{0,N}}{\sqrt{2}}$. 

In the small-$m$ limit (for $m=1$ in particular), 
we find $(\Delta \theta)_{\mathrm{QCR}} \leq 1/\langle \hat N \rangle$, 
violating Eq.~(\ref{HLtmsq}) \cite{GaoJPA2012}. 
The apparent contradiction between our results and Ref.~\cite{AnisimovPRL10} is solved by noticing that 
Eq.~(\ref{CRtmsq}) is known to be saturable (by the maximum likelihood estimator) only in the 
large-$m$ limit. There is no guarantee (and not shown in \cite{AnisimovPRL10}) that 
an unbiased estimator saturating Eq.~(\ref{CRtmsq}) for small-$m$ values can be found.  
The results of our manuscript show that such an unbiased estimator cannot exist.


\section{Conclusions}
Phase estimation will likely become the first large scale 
technology where classical bounds are overcome by quantum means.
It is therefore an interesting problem to set the 
fundamental quantum bound (generally indicated as the Heisenberg limit) 
which will limit the sensitivity of future phase estimation experiments. 
In this manuscript we have set the Heisenberg limit, Eq.~(\ref{HL}), 
under relevant experimental conditions: 
fluctuating number of particles, 
absence of number coherence in the probe state and/or 
in the measurement strategy and unbiased estimations. 
In this case we have also demonstrated that particle entanglement 
(we have extended the concept of particle entanglement to the case of state 
with fluctuating number of particles) is necessary to overcome the 
classical  -- shot noise -- phase uncertainty.
If the probe state and the output measurement contain coherences between different
total number of particles, it is not possible to establish a relation between entanglement and phase sensitivity and  
the phase sensitivity bound Eq.~(\ref{HL})
can only be set in the central limit. \\
\\*
{\it Acknowledgements.} 
We thank R. Demkowicz-Dobrza{\'n}ski for discussions. 
L.P. acknowledges support by MIUR through FIRB Project No. RBFR08H058.
P.H. acknowledges the support of the ERC starting grant GEDENTQOPT and the EU project QUASAR.


\section*{APPENDIX}


\subsection*{Appendix A: general two-mode transformations}

It is possible to write the general transformation (\ref{ModeMatrix}) as the product of four matrices \cite{CamposPRA1989}:
\begin{widetext}
\be 
\mathbf{U}=
\left[
\begin{array}{cc}
 e^{-i\phi_0} & 0 \\
 0 & e^{-i\phi_0} \\ 
\end{array} 
\right] \times 
\left[
\begin{array}{cc}
 e^{-i\psi/2} & 0 \\
 0 & e^{i \psi/2} \\ 
\end{array} 
\right] \times 
\left[
\begin{array}{cc}
 \cos \frac{\vartheta}{2} & -\sin \frac{\vartheta}{2} \\
 \sin \frac{\vartheta}{2} & \cos \frac{\vartheta}{2} \\ 
\end{array} 
\right] \times 
\left[
\begin{array}{cc}
 e^{-i\phi/2} & 0 \\
 0 & e^{+i \phi/2} \\ 
\end{array} 
\right],
\ee
\end{widetext}
where $\phi_\tau = (\psi + \phi)/2$ and $\phi_\rho = (\psi - \phi)/2$.
Using the Jordan-Schwinger representation of angular momentum, 
we have \cite{YurkePRA1986},
\beq
\vect{\mathrm{U}}_x = \left[
\begin{array}{cc}
 \cos \frac{\vartheta}{2} & -i\sin \frac{\vartheta}{2} \\
 -i\sin \frac{\vartheta}{2} & \cos \frac{\vartheta}{2} \\ 
\end{array} 
\right]  \,\, & \leftrightarrow & \,\, \hat U_x = e^{-i \vartheta \hat J_x}, \label{BS} \\
\vect{\mathrm{U}}_y = \left[
\begin{array}{cc}
 \cos \frac{\vartheta}{2} & -\sin \frac{\vartheta}{2} \\
 \sin \frac{\vartheta}{2} & \cos \frac{\vartheta}{2} \\ 
\end{array} 
\right]  \,\, & \leftrightarrow & \,\, \hat U_y = e^{-i \vartheta \hat J_y}, \label{MZ} \\
\vect{\mathrm{U}}_z = \left[
\begin{array}{cc}
 e^{-i\phi/2} & 0 \\
 0 & e^{+i \phi/2} \\ 
\end{array} 
\right]  \,\, & \leftrightarrow & \,\, \hat U_z = e^{-i \phi \hat J_z}, \label{PS}
\eeq
and using $e^{i\phi_0 \hat{N}} \, \hat a \, e^{-i\phi_0 \hat{N}} = e^{-i \phi_0} \hat a$, 
Eq.~(\ref{ModeMatrix}) can be associated to
\be \label{opap}
\hat{\mathrm{U}}(\phi_0, \theta) = e^{-i\phi_0 \hat{N}} e^{-i \psi \hat{J}_{\vect{z}}} e^{-i \vartheta \hat{J}_{\vect{y}}}
e^{-i \phi \hat{J}_{\vect{z}}}.
\ee
By using the Euler-Rodrigues formula, 
Eq.~(\ref{opap}) can be rewritten as 
\be
\hat{\mathrm{U}}(\phi_0, \theta) = e^{-i\phi_0 \hat{N}} e^{-i\theta \hat{J}_{\vect{n}}},
\ee
where
\be
\cos \frac{\theta}{2} = \cos\frac{\vartheta}{2} \cos \frac{\phi+\psi}{2},
\ee
and  $\hat J_{\vect{n}} = \alpha \hat J_{\vect{x}} + \beta \hat J_{\vect{y}} + \gamma \hat J_{\vect{z}}$, with
\beq
\alpha &=& \frac{\sin \frac{\vartheta}{2} \sin \frac{\phi-\psi}{2}}
{\sqrt{1 - \cos^2\frac{\vartheta}{2} \cos^2 \frac{\phi+\psi}{2}}}, \\
\beta &=& \frac{\sin \frac{\vartheta}{2} \cos \frac{\phi-\psi}{2}}
{\sqrt{1 - \cos^2\frac{\vartheta}{2} \cos^2 \frac{\phi+\psi}{2}}}, \\
\gamma &=& \frac{\cos \frac{\vartheta}{2} \sin \frac{\phi+\psi}{2}}
{\sqrt{1 - \cos^2\frac{\vartheta}{2} \cos^2 \frac{\phi+\psi}{2}}}. 
\eeq
This encompasses, for instance, the beam splitter [Eq.~(\ref{BS}), for $\phi=\pi/2$, $\psi = -\pi/2$ and $\vartheta=\theta$], 
Mach-Zehnder [Eq.~(\ref{MZ}), for $\phi=\psi=0$ and $\vartheta=\theta$] and 
phase shift [Eq.~(\ref{PS}), for $\psi=\vartheta=0$ and $\phi=\theta$] transformations.


\subsection*{Appendix B: derivation of Eq. (\ref{Peps})}

{\it States without number coherences.}
The incoherent probe Eq.~(\ref{eq:incoherent}) transforms according to Eq.~(\ref{ModeOperator}) as
\beq \label{rhoincev}
\hat\rho_\mathrm{out}(\phi_0,\theta) 
&=& \sum_{N=0}^{+\infty} Q_N \, \hat{\mathrm{U}}(\phi_0,\theta) \, \hat \rho^{(N)} \, \hat{\mathrm{U}}(\phi_0,\theta)^{\dag} \nonumber \\
&=& \sum_{N=0}^{+\infty} Q_N \, e^{-i \theta \hat J_{\vect{n}}} \hat \rho^{(N)} e^{+i \theta \hat J_{\vect{n}}},
\eeq
as a consequence of $[\hat \rho^{(N)}, \hat N]=0$.
Equation (\ref{rhoincev}) is a function of $\theta$ and shows that only SU(2) 
transformations, $e^{-i \theta \hat J_{\vect{n}}}$, are relevant for states without number coherence.
Equation~(\ref{Peps}) follows from Eq.~(\ref{rhoincev}), 
independently from the presence of number coherences in the POVM.

{\it POVMs without number coherence.}
For the case of states with coherences, Eq.~(\ref{eq:coherent}), 
and POVM without number coherences, Eq.~(\ref{NM_POVM}), we have
\beq 
P(\varepsilon\vert \theta) 
&=& \sum_{N} \tr\Big[ \hat \pi_N \, \hat E_N(\varepsilon) \, \hat \pi_N \, \hat{\mathrm{U}}(\phi_0,\theta) \, \hat \rho_{\rm coh} \, \hat{\mathrm{U}}(\phi_0,\theta)^{\dag} \Big] \nonumber\\
&=& \sum_{N} \tr\Big[ \hat E_N(\varepsilon) \, \hat{\mathrm{U}}(\phi_0,\theta) \, \hat \pi_N \hat \rho_{\rm coh}  \hat \pi_N \, \hat{\mathrm{U}}(\phi_0,\theta)^{\dag} \Big] \nonumber\\ 
&=& \sum_{N} Q_N \tr\Big[ \hat E_N(\varepsilon) \hat{\mathrm{U}}(\phi_0,\theta) \hat \rho^{(N)} \hat{\mathrm{U}}(\phi_0,\theta)^{\dag} \Big] \nonumber \\
&=& \sum_{N} Q_N \tr\Big[ \hat E_N(\varepsilon) e^{-i \theta \hat J_{\vect{n}}^{(N)}} \hat \rho^{(N)} e^{+i \theta \hat J_{\vect{n}}^{(N)}} \Big] \nonumber \\
&=& \sum_{N} Q_N P(\varepsilon\vert N, \theta), \label{Plambda}
\eeq
where
$P(\varepsilon\vert N, \theta) = \tr[ \hat E_N(\varepsilon) e^{-i \theta \hat J_{\vect{n}}^{(N)}} \hat \rho^{(N)} e^{+i \theta \hat J_{\vect{n}}^{(N)}}]$.
To derive this result we have used the commutation relation $[\hat{\mathrm{U}},\hat{\pi}_N]=0$ and 
the $\hat \pi_N \hat \rho_{\rm coh}  \hat \pi_N = Q_N \hat \rho^{(N)}$, where
$\hat \rho^{(N)}$ is a density matrix defined on the fixed-$N$ subspace.  
We have also used 
$\hat \pi_N \hat{\mathrm{U}}(\phi_0,\theta) \hat \pi_N = 
e^{-i \phi_0 N} e^{-i\theta \hat J_{\vect{n}}^{(N)}}$ due to the fact that 
$\hat J_{\vect{n}} = \oplus_{N} \hat J_{\vect{n}}^{(N)}$, 
where $\hat J_{\vect{n}}^{(N)}$ acts on the fixed-$N$ subspace.   
When POVM as in Eq.~(\ref{NM_POVM}) are used, we can therefore conclude, from Eq.~(\ref{Plambda}), that
only SU(2) transformations are relevant and number coherences in the probe state do not play any role.



\begin{thebibliography}{100}
\bibitem{nota:commutatorN} 
Let us consider a generic density matrix $\hat{\rho}$ as in Eq. (\ref{eq:coherent}) and recall that
$\ket{\psi_{N,k}}$ is a state of $N$ particles [it fulfills $\hat N \ket{\psi_{N,k}} = N \ket{\psi_{N,k}}$].
The commutator $[\hat \rho, \hat N] = \sum_k p_k \sum_{N,N'} \sqrt{Q_{N,k} Q_{N',k}^*} (N'-N) \ket{\psi_{N,k}} \bra{\psi_{N',k}}$ vanishes if and only if $\sqrt{Q_{N,k} Q_{N',k}^*} = \vert Q_{N,k} \vert \, \delta_{N,N'}$,
{\it i.e.} if and only if $\hat{\rho}$ can be written as in Eq.~(\ref{eq:incoherent}), 
with $Q_N \hat{\rho}^{(N)} = \sum_k p_k  \vert Q_{N,k} \vert \, \ket{\psi_{N,k}} \bra{\psi_{N,k}} 
= \hat{\pi}_N \hat \rho \hat{\pi}_N$.
\bibitem{nota:commutatorE}
Since $\sum_N \hat \pi_N = \Eins$, we can rewrite the POVM $\hat E(\varepsilon)$ as
$\hat E(\varepsilon) = \sum_{N,N'} \hat \pi_N \hat E(\varepsilon) \hat \pi_{N'}$.
Using $\hat N \hat \pi_N  = N \hat \pi_N$, we obtain 
$[\hat E(\varepsilon),\hat N] = \sum_{N,N'}  (N'-N) \hat \pi_N \hat E(\varepsilon) \hat \pi_{N'}$.  
This vanishes if and only if
$\hat \pi_N \hat E(\varepsilon) \hat \pi_{N'} = \delta_{N,N'}\hat \pi_N \hat E(\varepsilon) \hat \pi_{N}$,  
{\it i.e.} if and only if $\hat E(\varepsilon)$ has the form of Eq.~(\ref{NM_POVM}).
\bibitem{note:saturation} The inequality (\ref{CSCR}) is 
saturated if and only if 
\be
\vect{v}^{\top} \big[ \vect{\Theta}(\vect{\varepsilon}) - \vect{\bar{\Theta}} \big] 
= h(\vect{\theta}) \Big[ \frac{\partial}{\partial \vect{\theta}} \log \mathcal{L}(\vect{\varepsilon}|\vect{\theta}) \Big]^{\top} \vect{u},
\quad \forall \vect{\varepsilon}, \vect{\theta},
\ee 
where the function $h(\vect{\theta})$ does not depend on $\vect{\varepsilon}$.
\bibitem{nota_invposmat} We use the well known result that
every positive definite matrix is invertible and its inverse 
is also positive definite. In particular, if $\mathbf{A} \geq \mathbf{B} > 0$, 
then $\mathbf{B}^{-1} \geq \mathbf{A}^{-1} > 0$.
\end{thebibliography}
\end{document}